\documentclass[a4,11pt]{article}

\usepackage{amssymb}
\usepackage{amsmath}
\usepackage{amsfonts}
\usepackage{epsfig}
\usepackage{tabularx}
\usepackage[dvips]{color}
\usepackage{cite}

\newcommand{\be}{\begin{eqnarray}}
\newcommand{\ee}{\end{eqnarray}}

\newtheorem{theorem}{Theorem}[section]

\newtheorem{proposition}{Proposition}[section]
\newtheorem{lemma}{Lemma}[section]

\newtheorem{definition}{Definition}[section]
\newtheorem{remark}{Remark}[section]

\usepackage[subrefformat=parens]{subcaption}

\makeatletter
\def\Hline{%
\noalign{\ifnum0=`}\fi\hrule \@height 0.8pt \futurelet
\reserved@a\@xhline}
\makeatother

\usepackage[top=20truemm,bottom=20truemm,left=15truemm,right=15truemm]{geometry}

\begin{document} 

\begin{flushright}
\end{flushright}

\vspace{3.5cm}

\begin{center}
{\Large {\bf Stability of hypersurfaces with constant mean curvature \\ trapped between two parallel hyperplanes}}\\
\vspace{1.5cm}
{\bf Miyuki Koiso$^{a,1}$  and  Umpei Miyamoto$^{b,2}$} \\
\vspace{.5cm}
$^a$Institute of Mathematics for Industry, Kyushu University, 744 Motooka Nishi-ku, Fukuoka 819-0395, Japan\\
$^b$Research and Education Center for Comprehensive Science, Akita Prefectural University, 84-4 Tsuchiya Aza Ebinokuchi, Yurihonjo, Akita 015-0055, Japan\\
\vspace{.4cm}
$^1$koiso@imi.kyushu-u.ac.jp\\
$^2$umpei@akita-pu.ac.jp\\
\end{center}

\vspace{1cm}

\abstract{Static equilibrium configurations of continua supported by surface tension are given by constant mean curvature (CMC) surfaces which are critical points of a variational problem to extremize the area while keeping the volume fixed. CMC surfaces are used as mathematical models of a variety of continua, such as tiny liquid drops, stars, and nuclei, to play important roles in both mathematics and physics. Therefore, the geometry of CMC surfaces and their properties such as stability are of special importance in differential geometry and in a variety of physical sciences. In this paper we examine the stability of CMC hypersurfaces in arbitrary dimensions, possibly having boundaries on two parallel hyperplanes, by investigating the second variation of the area. We determine the stability of non-uniform liquid bridges or unduloids for the first time in all dimensions and all parameter (the ratio of the neck radius to bulge radius) regimes. The analysis is assisted by numerical computations.}

\newpage

\tableofcontents

\section{Introduction}\label{sec1}

A static equilibrium configuration of a continuum supported by surface tension is known to be well approximated by a 
constant-mean-curvature (CMC) surface, which extremizes the surface area for given volume and boundary conditions. CMC surfaces are used as mathematical models of a variety of continua, such as liquid drops, stars, and nuclei, to play important roles in both mathematics and physics~\cite{Kenmotsu, Capillarity}.

\begin{figure}[ht]
\centering
\includegraphics[width=4cm]{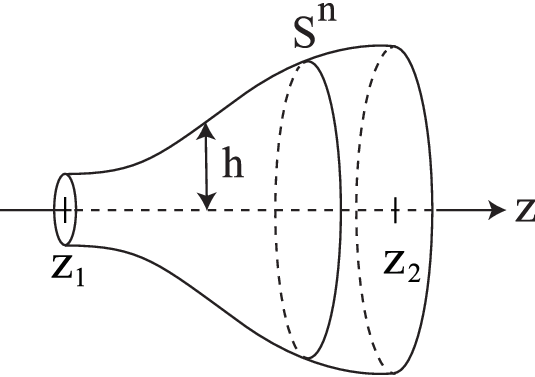}
\hspace{5mm}
\includegraphics[width=5cm]{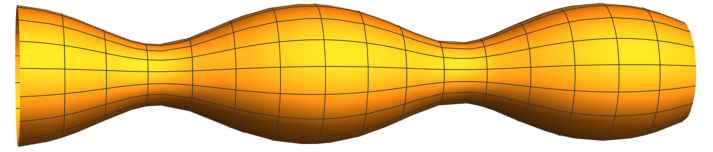}
\caption{{\it Left}: An axially-symmetric hypersurface in ${\mathbb R}^{n+2}$ with the $z$-axis as the axis of rotation. Its profile curve is given by a function $h=h(z)$ that represents the distance from the $z$-axis. {\it Right}: A part of an unduloid.}
\label{fig:hz}
\end{figure}

In the three-dimensional Euclidean space ${\mathbb R}^3$, 
one of fundamental problems regarding CMC surfaces is to find stable CMC surfaces which possibly have boundaries on given two parallel planes $\Pi_1$ and $\Pi_2$. Here, a CMC surface is said to be stable if the second variation of the area for any volume-preserving variation is non-negative. 
Since only stable surfaces are stably realized in natural phenomena, it is important to judge the stability for a CMC surface, which is difficult in general. 
Allowing no self-intersections of surface, it is shown that equilibrium surfaces contained in the region bounded by $\Pi_1$ and $\Pi_2$ are axially-symmetric CMC surfaces with the straight line perpendicular to $\Pi_i$ as its rotation axis \cite{K1986} (which will be called the $z$-axis throughout this paper), and they make contact angles $\pi/2$ with $\Pi_i \; (i=1,2)$\cite[Sect.1.6]{Finn}. Hence these surfaces are spheres, hemispheres, parts of cylinders and unduloids~\cite{Delaunay} (Fig. \ref{fig:hz}). Among them, only spheres, hemispheres, and thick cylinders are stable. Thin cylinders and unduloids are unstable CMC surfaces, {\it i.e.}, they extremize the area but do not minimize it for a given volume~\cite{Athanassennas,Vogel}.\footnote{In this paper, we are concerned with only the stability of a half period of unduloid ${\cal U}$ (from a neck to the next bulge or from a bulge to the next neck) since $m \times {\cal U} \; (m \geq 2)$ is always unstable. Note that if a half period of an unduloid is stable (resp.\ unstable) between two parallel hyperplanes in ${\mathbb R}^{n+2} \; (n \in {\mathbb N})$, one period of the unduloid is stable (resp.\ unstable) in ${\mathbb R}^{n+1} \times S^1$, and {\it vice versa}. This is proved in Appendix~\ref{period}.} The instability of thin cylinder is known as the Plateau-Rayleigh instability~\cite{Plateau, Rayleigh} in fluid mechanics. 

An interesting non-trivial aspect of the above problem is that the stability of unduloids depends on the dimension. 
There is a higher-dimensional counterpart of the unduloid
that is an axially-symmetric CMC hypersurface in ${\mathbb R}^{n+2}$ ($n=1, 2, \cdots$) which is periodic and has no self-intersection. 
We call it also an unduloid. 
A half period of an unduloid from a neck to the next bulge (such as the left figure in Fig. \ref{fig:hz}) satisfies the boundary condition. 
We define its non-uniformness parameter as 
$s := 1-(h_{\rm min}/h_{\rm max}) \in (0,1)$, 
where $h_{\rm min}$ and $ h_{\rm max} $ denote the radii of the unduloid at the neck and bulge, respectively. 
Then, 
for any negative number $H$ and $s \in (0, 1)$, up to rigid motion in ${\mathbb R}^{n+2}$, there exists exactly one unduloid with mean curvature $H$ with respect to the outward-pointing unit normal and having non-uniformness $s$. 
Denote by ${\cal U}={\cal U}(H, s)$ a half period of such unduloid. 
Then, for the Gromov-Hausdorff distance, $\lim_{s\to 1-0}{\cal U}(H, s)$ is a hemisphere with radius $1/\vert H \vert$, and $\lim_{s\to 0+0}{\cal U}(H, s)$ is a cylinder with radius $r=n/[(n+1)\vert H \vert]$ and height $L=(\sqrt{n}\pi)((n+1)\vert H \vert)^{-1}$. 
While the unduloids in higher dimensions were numerically obtained and their geometric quantities were computed~\cite{Miyamoto:2008rd, arXiv:0811.2305, Caldarelli:2008mv}, their stability has not been clarified completely so far. 
Let ${\cal U}$ be a half period of an unduloid. Then the following results on the stability are known \cite{Athanassennas,Vogel,PR,Li}.
\begin{itemize}
\item[{(i)}]
For any $n \geq 1$, if ${\cal U}$ is sufficiently close to a hemisphere, then ${\cal U}$ is unstable. 
\item[{(ii)}]
For $1 \!\leq \! n \!\leq 6$, ${\cal U}$ is unstable.
\item[{(iii)}]
For $7 \!\leq n\! \leq \! 9$ (resp. $n \!\geq \!10$), if ${\cal U}$ is sufficiently close to a cylinder, then ${\cal U}$ is unstable (resp.\ stable).
\item[{(iv)}]
For $n \geq 8$, there exists some ${\cal U}$ that is stable. 
\end{itemize}
In this paper, which corresponds to an extended version of a letter by the present authors \cite{KM2}, 
we comprehensively examine the stability of unduloids ${\cal U}$ in all dimensions and parameter regimes by investigating the second variation of area with the help of numerical computations. The results are summarized as statements (I)--(IV) in Sect.~\ref{sec:conclusion}.
A noteworthy result there is, roughly speaking, as follows: 
When $7 \leq n \leq 9$, if ${\cal U}$ is sufficiently close to either a cylinder or a hemisphere, then ${\cal U}$ is unstable, and moreover there exists stable ${\cal U}$.

Especially, the existence of stable unduloid for $n=7$ and instability of a half period of an unduloid close to a hemisphere for $7 \leq n \leq 9$ are found for the first time in this paper.

The geometric quantities of unduloid such as surface area, bulk volume, and mean curvature are obtained with the help of numerical integration. 
We will see that the stability is determined by the behaviors of these geometric quantities and stability criteria. 
There, besides the standard criteria for the stability, we use the bifurcation technique
 (see Sect. \ref{sec:sign}, \ref{sbif}) 
 developed in \cite{KPP2017} in order to judge the stability, which was not used in the papers mentioned above. 
It is remarkable 
that the regions of $s$ where the unduloid is stable (resp.\ unstable) completely coincide with those where the enclosed volume $V(s)$ is non-increasing (resp.\ increasing) for any $n$ (see Table~\ref{tbl:lambda2}, Sect. \ref{sec:conclusion}). 


Before starting analysis, let us mention that the higher-dimensional CMC hypersurfaces attract much attention in the study of general relativity, in particular, black holes. The black-hole counterparts of the cylinder and unduloid are called uniform black strings and non-uniform black strings respectively, and they exhibit various similarities with their counterparts~\cite{Sorkin:2004qq, Figueras:2012xj}. Furthermore, the `surface' of a black hole ({\it i.e.}, event horizon) was recently shown to indeed be approximated by a timelike CMC hypersurface in a large-dimension limit of general relativity~\cite{Emparan:2015hwa}. We will return this point in Sect.~\ref{sec:conclusion}.

The organization of this paper is as follows. We begin by calculating the variations of surface area and bulk volume for axially symmetric hypersurfaces in Sect.~\ref{sec:variation}. Then, an eigenvalue problem associated with the second variation of area is introduced in Sect.~\ref{sec:evp}. In Sect.~\ref{sec:criteria}, the stability criteria for unduloids is presented in terms of the eigenvalues, mean curvature, and volume. The stability of unduloids is examined in Sect.~\ref{sec:analysis}, using the criteria prepared in the previous section. Section~\ref{sec:conclusion} is devoted to summary and discussions. The proofs of mathematical propositions and the method to compute geometric quantities of unduloids are presented in Appendices~\ref{sec:math} and \ref{sec:num}, respectively. 

\section{Variation and eigenvalue problem}
\label{sec:setup}

\subsection{Area, volume, and their variations}
\label{sec:variation}

We consider axially symmetric hypersurfaces in the closed domain of ${\mathbb R}^{n+2} \; (n \in {\mathbb N})$ bounded by two hyperplanes $\Pi_1:=\{z=z_1\}$ and $\Pi_2:=\{z=z_2\}$.  The local radius of a hypersurface is represented by height function $h(z)$ (see Fig.~\ref{fig:hz})\footnote{Our main subject is to judge the stability of a half period of an unduloid. By using Schwarz symmetrization, we see that it is sufficient to study only axially symmetric variations. Namely, if the second variation of area is nonnegative for all axially symmetric volume-preserving variations of unduloid, then such an unduloid is stable. See Lemma~\ref{lsym2} in Appendix~\ref{sst} for a more general statement.}.

It is convenient to consider a one-parameter family of height function $h(z,\epsilon)$, where $\epsilon$ is a variation parameter. Then, the surface area and bulk volume of the axially symmetric object between $z=z_1$ and $z=z_2$ are
\begin{gather}
	A
	=
	a_n \int_{z_1}^{z_2} \sqrt{ 1+h_z(z,\epsilon)^2 } \; h(z,\epsilon)^n dz,
\label{area}
\\
	V
	=
	v_{n+1} \int_{z_1}^{z_2} h(z,\epsilon)^{n+1} dz.
\label{volume}
\end{gather}
Here, $a_n$ and $v_{n+1}$ are the volume of a unit $n$-sphere and that of a unit $(n+1)$-ball, respectively, given by
\begin{align}
	a_n
	= (n+1) v_{n+1}
	=  \frac{2\pi^{\frac{n+1}{2}}}{ \Gamma(  \frac{n+1}{2}) }.
\end{align}
A partial derivative is denoted by a subscript as $ h_z := \partial_z h $ hereafter. The mean curvature of the hypersurface is
\be \label{mc}
	H
	=
	\frac{1}{n+1} \Big[ \frac{ h_{zz} }{ (1+h_z^2)^{3/2} } - \frac{n}{h\sqrt{1+h_z^2}} \Big].
\ee
For a cylinder, hemisphere, and unduloid, $H$ takes a negative value in the present convention. 

The calculation of variations is equivalent to obtain the coefficients of the following expansion,
\begin{align}
\begin{split}
	X (\epsilon) = \sum_{\ell=0}^{\infty} \frac{ X_\ell }{ \ell ! } \epsilon^\ell, 
\;\;\;
	X =h , A , V , \; \mbox{or}\; H.
\end{split}
\end{align}
The coefficient of expansion is obtained by $X_\ell = \partial_\epsilon^\ell X \vert_{\epsilon=0} \; (\ell =0,1,2,\cdots)$.

The first variations of area and volume are easily obtained in terms of $h_0(z)$ and $h_1(z)$, 
\begin{gather}
	A_1
	=
	-(n+1) a_n \int_{z_1}^{z_2} H_0 h_0^n h_1 dz
	+
	a_n \Big[
		\frac{h_0^n h_{0z}  }{\sqrt{ 1+h_{0z}^2 }}h_1
	\Big]_{z_1}^{z_2},
\label{A1}
\\
	V_1
	=
	a_n \int_{z_1}^{z_2} h_0^n h_1 dz.
\label{V1}
\end{gather}
From Eqs.~\eqref{A1} and \eqref{V1}, one sees that the hypersurface which is the surface of revolution of $h_0$ is an equilibrium configuration or a critical point if and only if the following conditions hold,
\begin{gather}
	H_0(z)={\rm const.},
\label{eq:H0}
\\
 h_{0z}(z_1)=h_{0z}(z_2)=0.
\label{eq:h0}
\end{gather}
The CMC condition for the equilibrium configuration \eqref{eq:H0} corresponds to the Young-Laplace relation in fluid mechanics~\cite{LL}.

Now, let us focus on the volume-preserving variation ($ V_\epsilon \equiv 0 $) of CMC hypersurface, for which Eqs.~\eqref{eq:H0} and \eqref{eq:h0} hold. For such a variation, the first derivative of area can be written as
\begin{align}
	A_\epsilon
	&=
	A_\epsilon + (n+1)H_0 V_\epsilon
\label{A'2}
\\
	&=
	-(n+1)a_n \int_{z_1}^{z_2}
	\Big( H(z,\epsilon) - H_0 \Big) h^nh_\epsilon dz
	+
	a_n \left[
		\frac{  h^n h_z h_\epsilon }{ \sqrt{ 1+h_z^2 } } 
	\right]_{z_1}^{z_2}.
\label{A'3}
\end{align}
Then, the second derivative of area is
\begin{multline}
	A_{\epsilon\epsilon}
	=
	-(n+1) a_n \int_{z_1}^{z_2}
	\Big(
		H_\epsilon h^n h_\epsilon + n ( H-H_0 ) h^{n-1} h_\epsilon^2
		+ ( H-H_0 ) h^n h_{\epsilon \epsilon}
	\Big) dz
\\
	+
	a_n \left[ 
		\left(
			\frac{  h^n h_z h_\epsilon  }{ \sqrt{ 1+h_z^2 } } 
		\right)_{\!\!\! \epsilon} \; 
	\right]_{z_1}^{z_2}.
\label{A''}
\end{multline}
Using Eqs.~\eqref{eq:H0}, \eqref{eq:h0} and \eqref{A''}, one obtains the second variation of area in terms of $h_0$, $h_1$, and $H_1$,
\begin{align}
	A_2
	=
	-(n+1)a_n\int_{z_1}^{z_2}  H_1(z) h_0^n h_1  dz
	+
	a_n \left[ h_0^n h_1 h_{1z} \right]_{z_1}^{z_2}.
\label{A2_pre}
\end{align}
It is noted that $A_2$ is independent of $h_2$ due to the addition of term $ (n+1)H_0 V_\epsilon $ in Eq.~\eqref{A'2}.

The first variation of mean curvature $H_1$ in Eq.~\eqref{A2_pre} can be written as  
\begin{align}
	H_1(z)
	=
	\frac{1}{	(n+1)h_0^n} \mathcal{L} h_1,
\label{ell}
\end{align}
by defining the following linear operator
\begin{gather}
	\mathcal{L}
	:=
	\frac{d}{dz}\left(  \sigma(z) \frac{d}{dz} \right) + \frac{ nh_0^{n-2} }{ \sqrt{1+h_{0z}^2} },
\label{jacobi}
\\
	\sigma(z)
	:=
	\frac{ h_0^n }{ \left( 1+h_{0z}^2 \right)^{3/2} }.
\label{operator}
\end{gather}
Therefore, $A_2$ is written in a simple form,
\begin{align}
	A_2
	=
	-a_n\int_{z_1}^{z_2}
	h_1 \mathcal{L} h_1 dz
	+
	a_n \left[  h_1 \sigma(z) h_{1z} \right]_{z_1}^{z_2}.
\label{A2}
\end{align}

\subsection{Eigenvalue problem associated with second variation of area}
\label{sec:evp}

An equilibrium is defined to be stable if the second variation is non-negative for all volume-preserving variations. This condition is equivalent to $A_2  \geq 0$ for all variations satisfying $V_1 = 0$.

From this viewpoint, let us consider the following eigenvalue problem associated with $A_2$.
\begin{align}
\begin{split}
	{\cal L} \varphi_i (z) &= -\lambda_i \varphi_i (z),
\\
	\varphi_{iz}(z_1) & = \varphi_{iz}(z_2) = 0,
\end{split}
\label{evp}
\end{align}
where $i=1,2,3,\cdots$ labels the eigenvalue $\lambda_i$ and eigenfunction $\varphi_i (z)$.
Since ${\cal L}$ is a Sturm-Liouville operator, it is shown that $ \lambda_1 < \lambda_2 < \lambda_3 < \cdots $, and that $\varphi_i (z)$ has exactly $i-1$ zeros in $( z_1,z_2 )$.

The general variation of the height function is a linear combination of the eigenfunctions $ h_1(z) = \sum_{i=1}^\infty c_i \varphi_i (z)$, $c_i \in {\mathbb R}$.  Then, $A_2$ and $V_1$ are written in terms of $c_i$ and $\lambda_i$,
\begin{gather}
	A_2 = a_n \sum_{i=1}^\infty c_i^2 \lambda_i,
\label{A2b}
\\
	V_1 = a_n \sum_{i=1}^\infty c_i \int_{z_1}^{z_2} h_0^n \varphi_i dz,
\label{V1b}
\end{gather}
where the orthonormality $ \int_{z_1}^{z_2} \varphi_i \varphi_j dz = \delta_{ij} $ is assumed. 

From Eqs.~\eqref{A2b} and \eqref{V1b}, one sees that an equilibrium $h_0$ is stable if $\lambda_1 \geq 0$ since in such a case $A_2 > 0$ for all non-trivial ({\it i.e.}, $h_1 \not\equiv 0$) volume-preserving variations satisfying $V_1=0$. One the other hand, one sees that {\it an equilibrium $h_0$ is unstable if $\lambda_2$ is negative.} Namely,
\be
	\lambda_2 < 0 \;\;\; \Longrightarrow \;\;\; \mbox{unstable}
\label{cr0}
\ee
holds since in such a case $A_2 < 0$ for the volume-preserving variation given by
\begin{align}
	 c_1 = - \frac{ \int_{z_1}^{z_2} h_0^n \varphi_2 dz }{ \int_{z_1}^{z_2} h_0^n \varphi_1 dz } c_2,
\;\;\;
	c_2  \neq  0,
\;\;\;
	c_i = 0 \;\;\; (i \geq 3).
\end{align} 

For a uniform cylinder $h_0 \equiv r = {\rm const.}$, Eq.~\eqref{evp} is
\begin{align}
	\frac{ d^2 \varphi_i }{dz^2}  + \frac{ nr^{n-2} + \lambda_i^{\rm cyl} }{ r^n } \varphi_i = 0.
\label{eq:evp_cyl}
\end{align}
If one puts $z_1 = 0, z_2  = L \; (>0) $, the eigenvalue of a cylinder $\lambda_i^{\rm cyl}$ is obtained by solving Eq.~\eqref{eq:evp_cyl},
\be
	\lambda_i^{\rm cyl}
	=
	\left(
		\Big[ \frac{ (i-1) \pi r}{L} \Big]^2-n
	\right) r^{n-2},
\;\;\;
	i=1,2,3,\cdots.
\label{lambda^cyl}
\ee
From Eq.~\eqref{lambda^cyl}, one can see that if 
\begin{align}
	r < r_c := \frac{\sqrt{n} L}{\pi},
\end{align}
$\lambda_2^{\rm cyl} < 0$ holds and such a thin cylinder is unstable from criterion \eqref{cr0} (see also Refs.~\cite{Cardoso:2006ks, Miyamoto:2008uf} for a dynamical counterpart). More precisely, it is proved that the cylinder with radius $r$ and length $L$ is stable if and only if $r\ge r_c$ holds ({\it cf.}\ \cite{Li}). We call the cylinder with critical radius $r_c$ {\it a critical cylinder}.

The sphere $S^{n+1}$ and the hemisphere with a boundary in either $z=z_1$ or $z=z_2$ are stable because $S^{n+1}$ is the minimizer of area among all closed hypersurfaces enclosing the same volume.

\section{Stability criteria of unduloids}
\label{sec:criteria}

It is convenient to introduce a quantity parameterizing the family of unduloids. As such a quantity, we adopt the non-uniformness parameter 
\be
	s = 1-\frac{h_{\rm min}}{h_{\rm max}} \in (0,1),
\label{nonuni}
\ee
introduced in Sect.~\ref{sec1} where $h_{\rm min}$ and $ h_{\rm max} $ denote the radii of an unduloid at the neck and bulge, respectively. One can naturally assign $s=0$ and $s=1$ to the critical cylinder and the largest hemisphere, that fits the interval, respectively. In the rest of this paper, we denote the half period of unduloid itself, mean curvature, volume, and eigenvalue of such an unduloid by ${\cal U}(s)$, $H(s)$, $V(s)$, and $\lambda_i(s)$, respectively.\footnote{In this paper, we assume the continuity of $\lim_{s \to +0} X = X \vert_{s=0}$ and $\lim_{s \to 1-0} X = X\vert_{s=1}$ for $X=A, V$ and $H$.}

For ${\cal U}(s)$, one can show the negativity (resp.\ positivity) of $\lambda_1$ (resp.\ $\lambda_3$). Namely, the following holds,
\be
	\lambda_1(s) < 0 < \lambda_3(s), \;\;\; \forall s \in (0,1).
\label{lambda13}
\ee
See Appendix~\ref{sec:lambda13} for a proof.

In the rest of this section, we will introduce mathematical theories which play crucial roles in the stability analysis of Sect.~\ref{sec:analysis}. In Sect.~\ref{sec:sign}, we see how to determine the sign of $\lambda_2(s)$ from the behavior of $H(s)$. While $\lambda_2(s)<0$ immediately implies the instability of ${\cal U}(s)$ from \eqref{cr0}, another criterion is needed to determine the stability of ${\cal U}(s)$ when $\lambda_2(s) \geq 0$. Therefore, in Sect.~\ref{sec:positive}, we see how the behavior of $H(s)$ and $V(s)$ determines the stability when $\lambda_2(s) \geq 0$.

\begin{figure}[t]
\centering
\tabcolsep=5mm
		\begin{tabular}{ c c }
			\includegraphics[height=30mm]{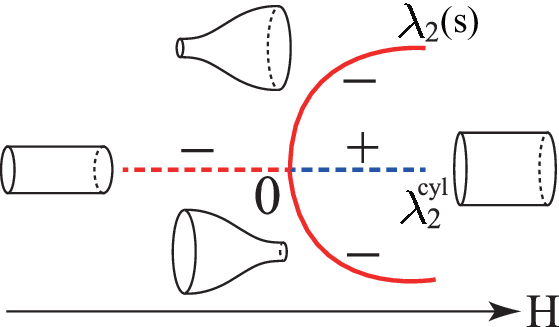} &
			\includegraphics[height=30mm]{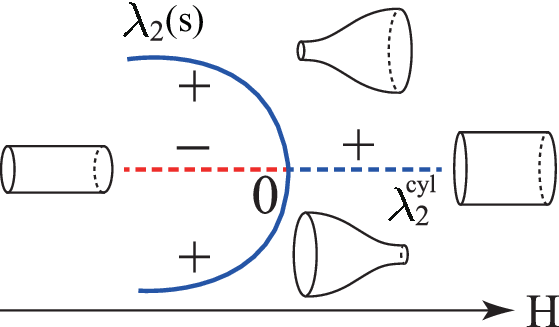} \\
			(a) & (b) \\
		\end{tabular}
\caption{Diagrams representing criterion \eqref{cr1}. If the mean curvature of unduloid is (a) larger (resp.\ (b) smaller) than that of the critical cylinder $H(s) > H(0)$ (resp.~$H(s) < H(0)$), $\lambda_2 $ is negative (resp.\ positive). Note that $H$ takes negative values in the present convention.}
\label{fig:bif}
\end{figure}

\subsection{Sign of second eigenvalue $\lambda_2$}
\label{sec:sign}

From Eq.~\eqref{lambda^cyl}, one can see that the second eigenvalue of cylinder $\lambda_2^{\rm cyl}$ increases and changes sign from negative to positive as radius $r$ increases. From the point where $\lambda_2(0)=0$, two branches of unduloid\footnote{The half period of unduloid with a neck at $z_1$ and one with a bulge at $z_1$ are distinguished in the current bifurcation theory, although their physical properties are identical.} emanate (see Fig.~\ref{fig:bif}). For these branches of unduloids bifurcating from the critical cylinder, the sign of $\lambda_2 (s)$ is determined by the relative value of the mean curvature to that of the critical cylinder. Namely, {\it if the mean curvature of the emanating unduloid $H(s)$ is larger (resp.\ smaller) than that of the critical cylinder $H(0)$, the second eigenvalue of the unduloid $\lambda_2(s)$ is negative (resp.\ positive),}
\begin{align}
	H(s) \gtrless H(0) \;\;\; \Longrightarrow \;\;\; \lambda_2 (s) \lessgtr 0.
\label{cr1}
\end{align}
This statement is derived from Theorem \ref{bifurcation-stabilityII} in Appendix~\ref{sbif}. 
The criterion is visualized in Fig.~\ref{fig:bif}. This criterion can be utilized to determine the sign of $\lambda_2(s)$ for ${\cal U}(s)$ close to the cylinder $0<s \ll 1$.

In order to know when $\lambda_2(s)$ changes sign, the following criteria are quite useful.
\begin{align}
	\mbox{$H' \neq 0$ at $s$}\;
	& \; \Longrightarrow \;\;
	\mbox{$\lambda_2$ does not change sign at $s$},
\label{cr21}
\\
	\mbox{$H'=0$ \& $H''\ne 0$ \& $V'\ne 0$ at $s$}\;
	& \; \Longrightarrow \;\;
	\mbox{$\lambda_2=0$ \& $\lambda_2$ changes sign at $s$}.
\label{cr22}
\end{align}
Their proofs are presented in Appendix~\ref{sign}.  What criteria \eqref{cr21} and \eqref{cr22} mean is that, under the assumption that $H''(s)\ne 0$ and $V'(s)\ne 0$, $\lambda_2(s)$ changes sign when $H'(s)$ does. Although the possibility that $\lambda_2(s) $ vanishes even when $H'(s) \neq 0$ is not excluded by \eqref{cr21}, it can be proved that 
\begin{align}
	H'(s) \neq 0 \;\;\; \Longrightarrow \;\;\; \lambda_2(s) \neq 0.
\label{cr23} 
\end{align}
See Prop.\ \ref{l2} in Appendix \ref{vanishment}. Thus, once the sign of $\lambda_2(s)$ near $s=0$ (the critical cylinder) is determined by \eqref{cr1}, the sign of $\lambda_2$ in the full range of $s$ is known by investigating $H''(s)$ and $V'(s)$ at zeros of $H'(s)$.

\subsection{Criteria when $\lambda_2 \geq 0$}
\label{sec:positive}

While $\lambda_2(s)<0$ immediately implies that an unduloid ${\cal U}(s)$ is unstable from \eqref{cr0}, another criteria is necessary to determine the stability of ${\cal U}(s)$ when $\lambda_2 (s) \geq  0$. 
From the criteria for the stability given by Lemma \ref{slem1} in Appendix~\ref{sst}, we have the following observations.

{\it When $\lambda_2(s) = H'(s)= 0$ and $V'(s) \ne 0$ hold, ${\cal U}(s)$ is unstable}. Namely, the following holds,
\be
	\mbox{ $\lambda_2(s)=H'(s)= 0$ \& $V'(s) \neq 0$}
\;\;\; \Longrightarrow \;\;\;
	\mbox{unstable. }
\label{cr3}
\ee
When $\lambda_2(s) > 0$, the stability is related to the increase and the decrease of the mean curvature and volume. Namely, {\it ${\cal U}(s)$ with $\lambda_2(s) > 0$ is unstable (resp.\ stable) if $H'(s)V'(s)$ is negative (resp.\ non-negative)}. 
\begin{align}
\begin{split}
	\lambda_2(s) > 0
	\;\;\;
	\&
	\;\;\;
	\begin{cases}
	H'(s) V'(s) < 0 \\
	H'(s) V'(s) \geq 0
	\end{cases}
\;\;\; \Longrightarrow \;\;\;\;
	\begin{cases}
	\mbox{unstable}\\
	\mbox{stable}
	\end{cases}.
\end{split}
\label{cr4}
\end{align}
We will utilize criteria~\eqref{cr3} and \eqref{cr4} to determine the stability of ${\cal U}(s)$ whose $\lambda_2(s)$ is non-negative.

\subsection{Comment: No iteration is needed}
\label{sec:notes}

In the next section, we numerically obtain the mean curvature and volume for each ${\cal U}(s)$. Before starting such an analysis, let us see that obtaining $H(s)$, $V(s)$, and their derivatives numerically is a much simpler task than solving eigenvalue problem \eqref{evp}.

$H(s)$ and $V(s)$ can be computed by just obtaining  the `background' solution $h_0(z)$. The function $h_0(z)$ is obtained by solving   $H_0(z)={\rm const.}$ with boundary conditions $h_{0z}(z_1)=h_{0z}(z_2)=0$. At a first glance, this problem seems to be a two-boundary problem requiring an iterative integration. By reducing $H_0(z)={\rm const.}$, which is a second-order ODE (ordinary differential equation), to an equivalent potential problem (a first-order ODE) and introducing an appropriate parameterization, however, no iteration turns out to be needed and the geometric quantities of unduloids, $H(s)$ and $V(s)$, are obtained by just estimating several improper integrals numerically (see Appendix~\ref{sec:num}).  

On the other hand, the eigenvalue equation \eqref{evp} is essentially a two-boundary problem requiring an iteration procedure such as the shooting method~\cite{recipe}. Furthermore, one has to numerically solve the ``perturbation equation'' \eqref{evp} for $\psi_2(z)$ and $\lambda_2$ on the numerical background $h_0(z)$, which is a part of operator ${\mathcal L}$ in Eq.~\eqref{jacobi}.

Thus, it is stressed here that the stability criteria presented in Sects.~\ref{sec:sign} and \ref{sec:positive} are not only easy to use but also enormously reduce the amount of numerical computations required in the analysis. 
This demonstrates the merit of adopting the geometric variational method throughout in our analysis, rather than ordinary mode-decomposition methods which are the standard for stability analysis in physics.

\section{Stability of unduloids in ${\mathbb R}^{n+2}$ ($ n \in {\mathbb N} $)}
\label{sec:analysis}

What are needs to do in order to examine the stability of all unduloids is to obtain the height function $h_0(z)$ corresponding to the half period of unduloid by numerically integrating the ODE $H_0(z)={\rm const}.$ with boundary conditions $h_{0z}(z_1)=h_{0z}(z_2)=0$, while taking the dimension $n \in {\mathbb N}$ and non-uniformness $s \in (0,1)$ as free parameters. Then, one can estimate the mean curvature $H$ and volume $V$ as functions of $s$ for each $n$.\footnote{In fact, the ODE $H_0(z)={\rm const}.$ can be reduced to an equivalent first-order ODE and the geometric quantities (area, volume, and mean curvature) of unduloids are then written as improper integrals, which are functions of only $s$ (after fixing the period of unduloid) for each $n$. Therefore, what we need to do is to estimate such improper integrals accurately. See Appendix~\ref{sec:num} for details.} Finally, utilizing the stability criteria \eqref{cr0} and \eqref{cr1}--\eqref{cr4}, one can determine the stability of every unduloid.



In the rest of this section, we will clarify the stability of unduloids in all dimensions and parameter regimes of $s$. According to the behaviors of geometric quantities, we classify the dimensions into four classes, A ($ 1\leq n \leq 6$), B ($n=7$), C ($ 8 \leq n \leq 9$), and D ($ n \geq 10 $), and examine the stability separately. Qualitative features of diagrams and stability structures are common in each class. The results in a final form are summarized as four statements (I)--(IV) in Sect.~\ref{sec:conclusion}.

\begin{figure}
\centering
\tabcolsep=2mm
		\begin{tabular}{ c c }
			\multicolumn{2}{c}{$n=6 \in \mbox{Class A}$}\\
			\includegraphics[height=30mm]{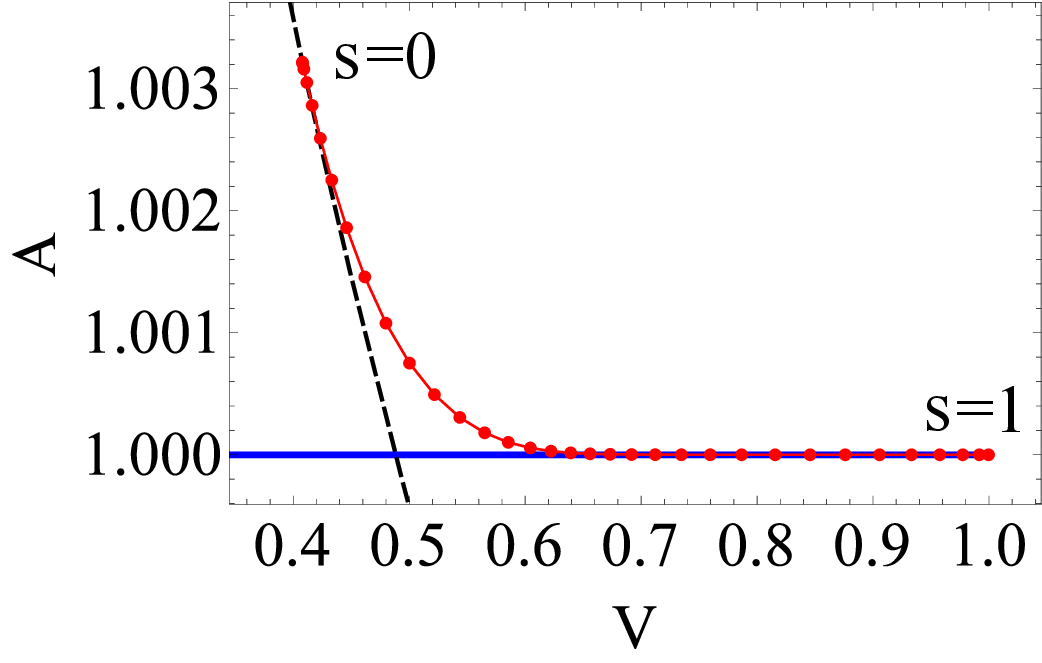} &
			\includegraphics[height=30mm]{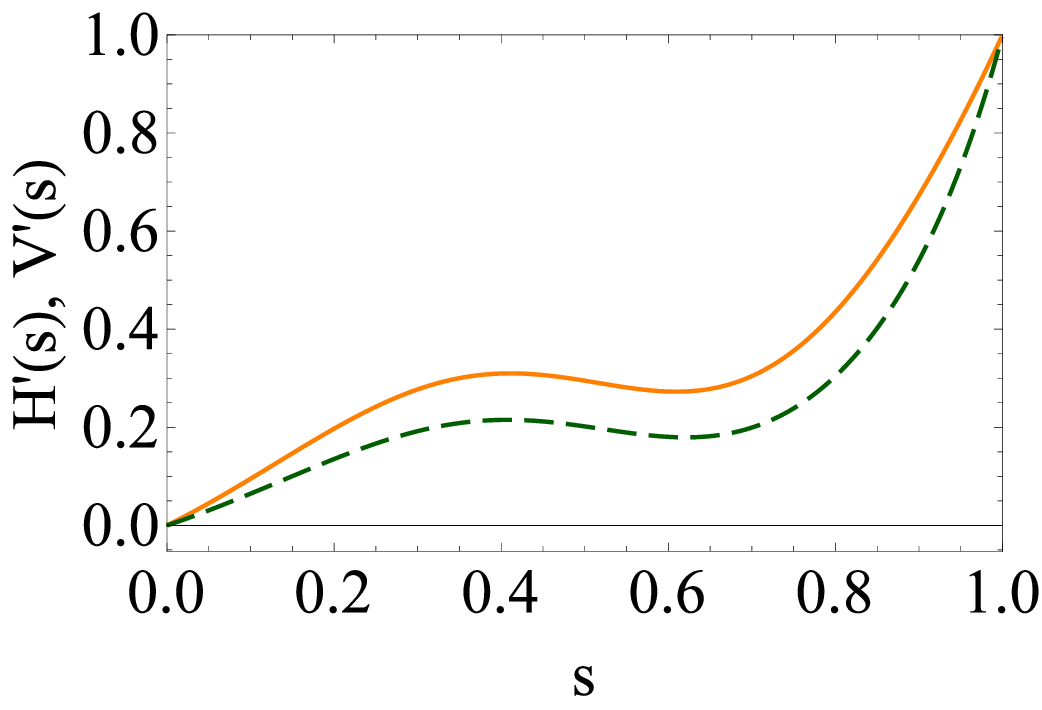} \\
			\multicolumn{2}{c}{$n=7 \in \mbox{Class B}$}\\
			\includegraphics[height=30mm]{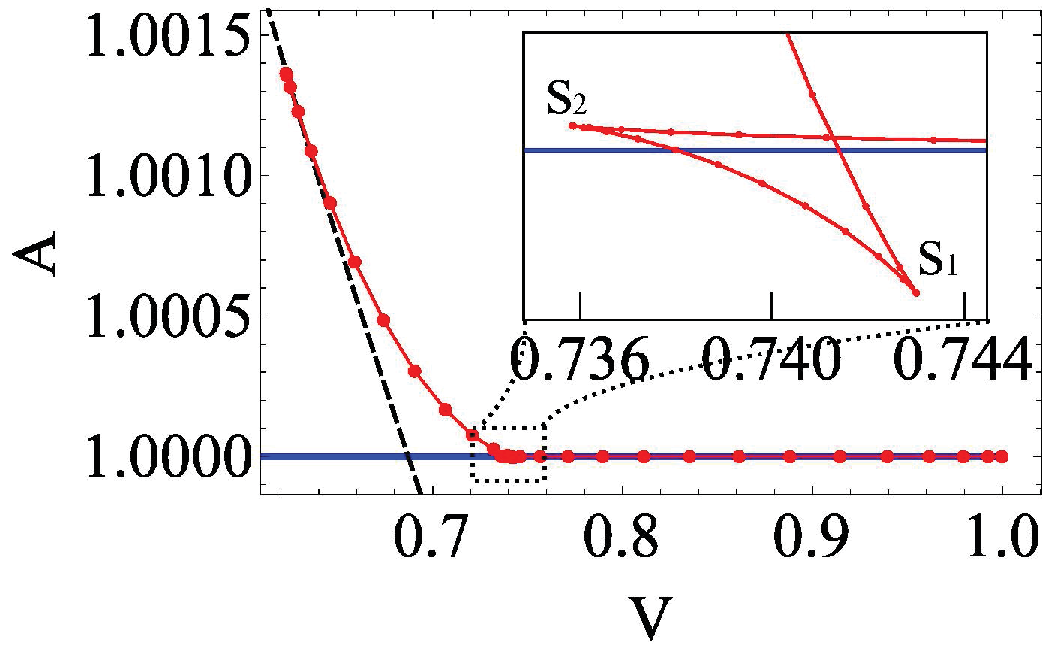} &
			\includegraphics[height=30mm]{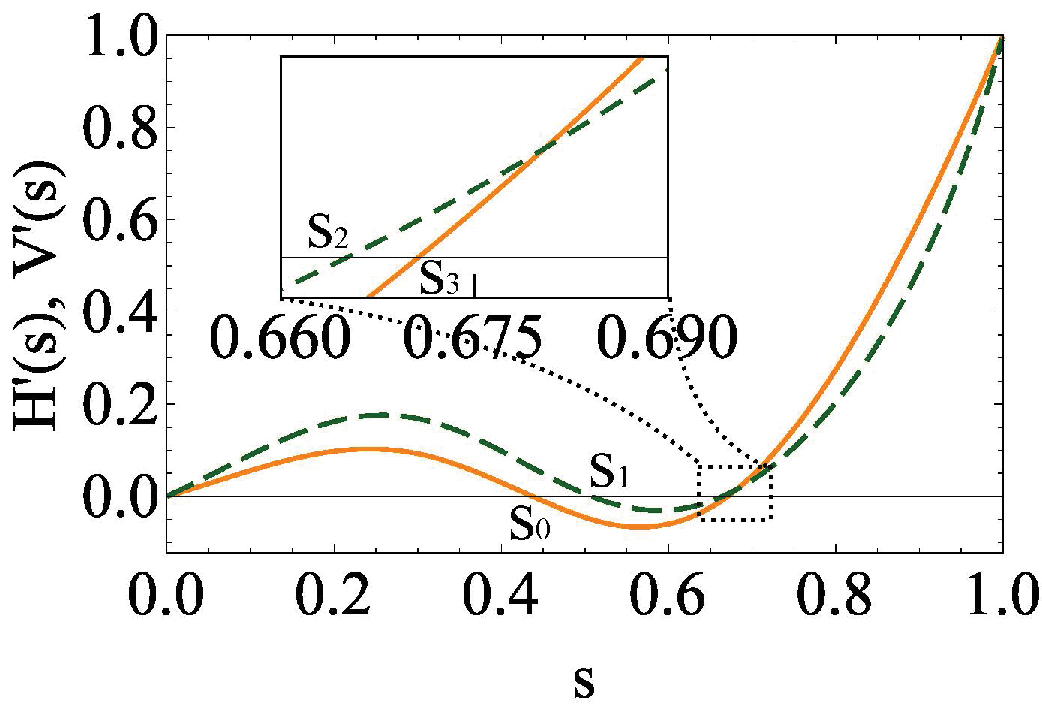} \\
			\multicolumn{2}{c}{$n=8 \in \mbox{Class C}$}\\
			\includegraphics[height=30mm]{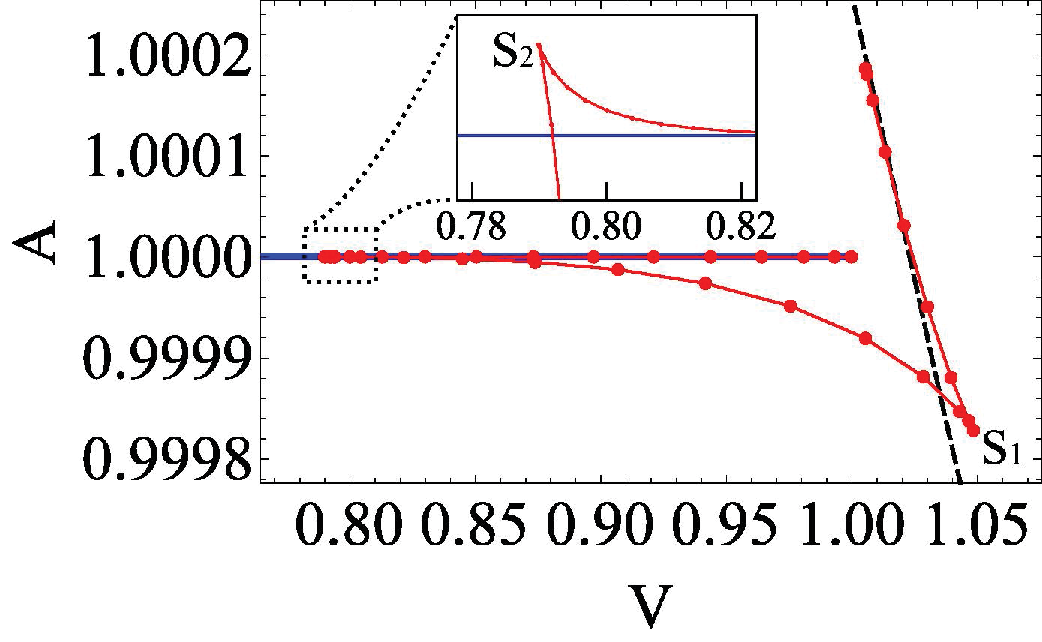} &
			\includegraphics[height=30mm]{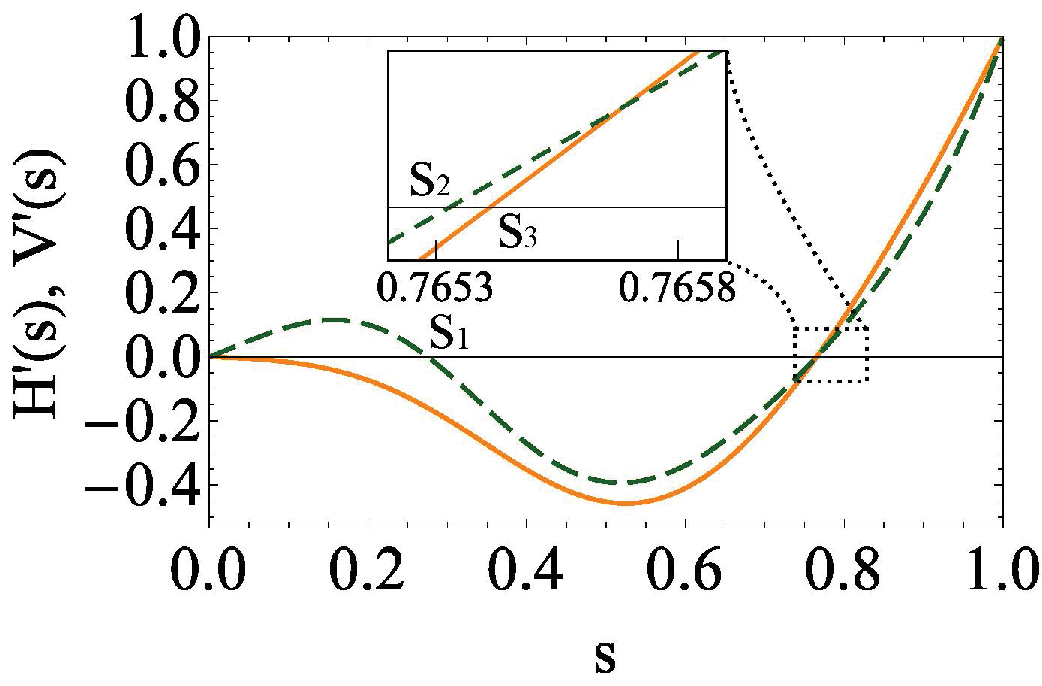} \\
			\multicolumn{2}{c}{$n=10 \in \mbox{Class D}$}\\
			\includegraphics[height=30mm]{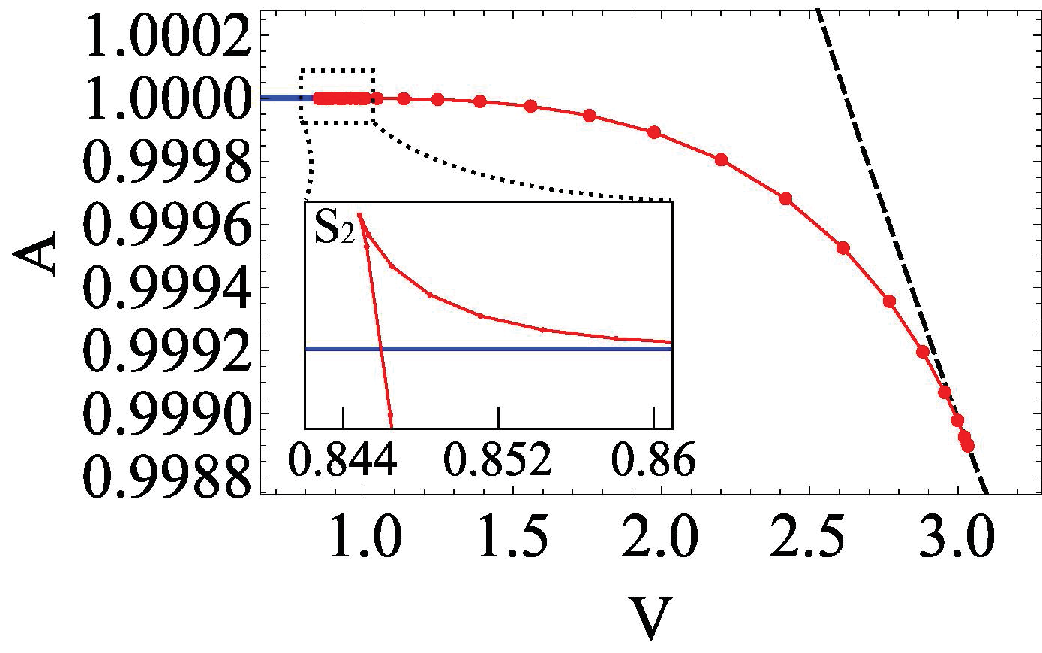} &
			\includegraphics[height=30mm]{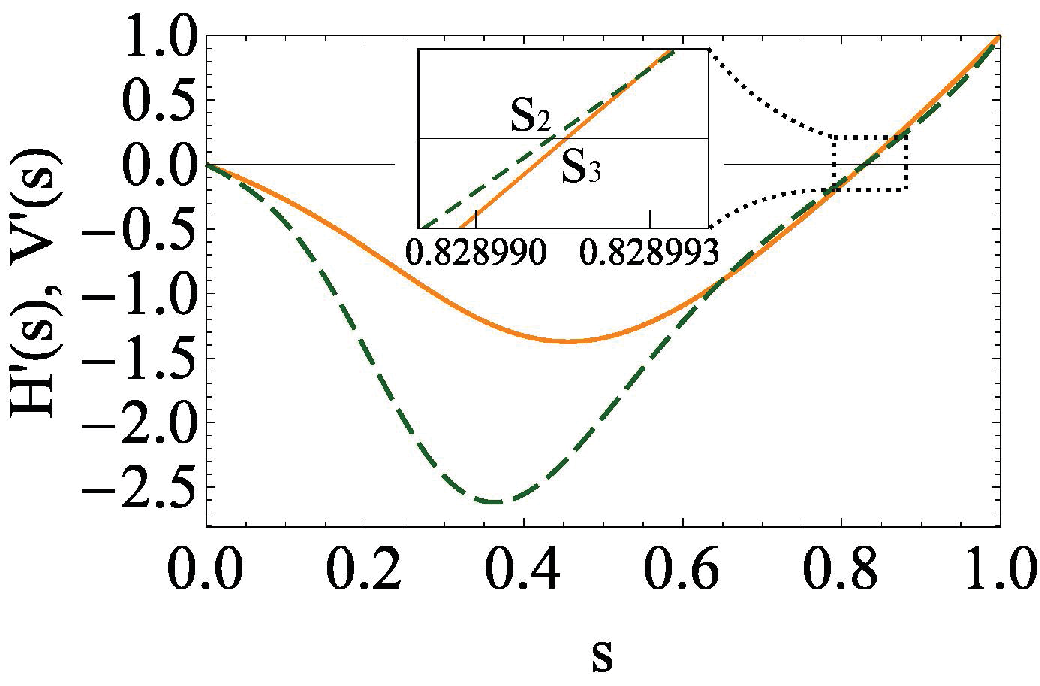} \\
		\end{tabular}
\caption{{\it Left}: Area-volume diagrams of the cylinder (dashed black line), hemisphere (thick blue line), and unduloid (red dots with solid line) for $n=6, 7, 8$, and $10$, from the top to the bottom. {\it Right}: $H'(s)$ (solid orange line) and $V'(s)$ (dashed green line) of unduloid ${\cal U}(s)$ for $n=6, 7, 8$, and $10$ from the top to the bottom.}
\label{fig:diagrams}
\end{figure}

The characteristic area-volume diagrams of the unduloid, cylinder, and hemisphere are shown in Fig.~\ref{fig:diagrams}. In addition, the numerical plots of $H'(s)$ and $V'(s)$, the derivatives of mean curvature and volume of unduloid ${\cal U}(s)$ with respect to $s$, are presented also in Fig.~\ref{fig:diagrams}.

The area in the area-volume diagram is normalized in such a way that the area of the hemisphere remains unity in all ranges of the volume. The volume is normalized in such a way that the volume of the largest hemisphere, which fits the interval $[z_1, z_2]$, is unity. In the plots of $H'(s)$ and $V'(s)$, $H'(s)$ and $V'(s)$ are normalized by $\lim_{s \to 1-0} \lvert H'(s) \rvert =1 $ and $\lim_{s \to 1-0} \lvert V'(s) \rvert =1$, respectively.

\begin{table}[ht]
\centering
\tabcolsep=3 mm

\caption{The sign of $H'(s), \lambda_2(s)$, and $V'(s)$ and the stability of unduloid ${\cal U}(s)$ in ${\mathbb R}^{n+2}$ as functions of the non-uniformness parameter $s \in (0,1)$ in Class A, B, C, and D.}
\label{tbl:lambda2}

\vspace{5pt}

  $\mbox{Class A}$\\
\begin{tabularx}{42mm}{c|  c     }
\Hline
$s$ &   \\
\hline
$H'(s)$ &  $+$   \\
\hline
$\lambda_2(s)$ &  $-$   \\
\hline
$V'(s)$ & +   \\
\hline
stability &  unstable     \\
\Hline
\end{tabularx}

\vspace{10pt}

$\mbox{Class B}$\\
\begin{tabularx}{100mm}{c|  c | c | c | c | c | c | c | c | c }
\Hline
$s$ & & $s_0$ & & $s_1$ & & $s_2$ & & $s_3$ &  \\
\hline
$H'(s)$ &  $+$  & 0 & \multicolumn{5}{c|}{$-$} & $0$ & $+$\\
\hline
$\lambda_2(s)$ &  $-$  & 0 & \multicolumn{5}{c|}{$+$} & $0$ & $-$\\
\hline
$V'(s)$ & \multicolumn{3}{c|}{$+$}  & 0 & $-$& 0 &  \multicolumn{3}{c}{$+$} \\
\hline
stability &  \multicolumn{3}{c|}{unstable}  &  \multicolumn{3}{c|}{stable} & \multicolumn{3}{c}{unstable} \\
\Hline
\end{tabularx}

\vspace{10pt}

$\mbox{Class C}$\\
\begin{tabularx}{100mm}{c|  c | c | c | c | c | c | c  }
\Hline
$s$ & & $s_1$ & & $s_2$ & & $s_3$ &    \\
\hline
$H'(s)$ &  \multicolumn{5}{c|}{$-$}  & 0 & $+$ \\
\hline
$\lambda_2(s)$ &  \multicolumn{5}{c|}{$+$}  & 0 & $-$\\
\hline
$V'(s)$ & $+$  & 0 & $-$& 0 &  \multicolumn{3}{c}{$+$} \\
\hline
stability &  unstable  &  \multicolumn{3}{c|}{stable} & \multicolumn{3}{c}{unstable} \\
\Hline
\end{tabularx}

\vspace{10pt}

$\mbox{Class D}$\\
\begin{tabularx}{65mm}{c|  c | c | c | c | c }
\Hline
$s$ & & $s_2$ & & $s_3$ & \\
\hline
$H'(s)$ &  \multicolumn{3}{c|}{$-$}  & 0 & $+$ \\
\hline
$\lambda_2(s)$ &  \multicolumn{3}{c|}{$+$}  & 0 & $-$\\
\hline
$V'(s)$ & $-$  & 0 & \multicolumn{3}{c}{$+$} \\
\hline
stability &  \multicolumn{2}{c|}{stable} & \multicolumn{3}{c}{unstable} \\
\Hline
\end{tabularx}

\end{table}

\subsection{Class A: $1 \leq n \leq 6$}
\label{sec:n6}

It is characteristic of this class that any unduloid has larger area than the cylinder and hemisphere with the same volume, and the area-volume curve of an unduloid has no cusp.

From Fig.~\ref{fig:diagrams}, one can see that $H'(s) > 0, \; \forall s \in (0,1)$. Therefore, the sign of $\lambda_2(s)$ is definite in the range of $s$ by criteria \eqref{cr21} and \eqref{cr23}. From the fact that $H'(s)>0$ in a vicinity of $s=0$, $H(s)>H(0)$ holds for the unduloid which just bifurcated from the critical cylinder, which implies $\lambda_2(s)<0$ near $s=0$ with criterion \eqref{cr1}  (see also Fig.~\ref{fig:bif}). Thus, $\lambda_2(s) < 0$ holds $\forall s \in (0,1)$, from which \eqref{cr0} implies all unduloids are unstable in this class. This is consistent with known results in the literature, {\it i.e.}, statement~(ii).

\subsection{Class B: $n=7$}
\label{sec:n7}

It is characteristic of this class that the area-volume curve of an unduloid has two cusps which form a swallowtail shape. We observe that both $H'(s)$ and $V'(s)$ have two simple zeros, which we denote by $s_k$ ($ k=0,1,2,3$) as
\begin{align}
\begin{split}
	H'(s_0)= V'(s_1) & =V'(s_2)=H'(s_3)=0,  \\
	0<s_0<s_1  &  <s_2<s_3<1.
\end{split}
\label{n7s}
\end{align}

From the behavior of $H'(s)$, one knows that $\lambda_2(s)$ vanishes and changes sign only at $s=s_0$ and $s=s_3$ with criteria \eqref{cr21}-\eqref{cr23}. From this fact and the behavior of $H'(s)$ with criterion \eqref{cr1}, one can see that $\lambda_2(s) < 0 $ (resp.\ $\lambda_2(s) \geq 0 $) for $s \in (0,s_0) \cup (s_3,1)$ (resp.\ $s \in [s_0,s_3]$). Therefore, ${\cal U}(s)$ for $s \in (0,s_0) \cup (s_3,1)$ is unstable with criterion \eqref{cr0}. Since $\lambda_2(s) \geq 0 $ for $s \in [s_0,s_3]$, we have to see also the behavior of $V'(s)$ in order to use criteria~\eqref{cr3} and \eqref{cr4}. From Fig.~\ref{fig:diagrams}, $V'(s)$ vanishes at neither $s=s_0$ nor $s=s_3$, which with \eqref{cr3} implies that ${\cal U}(s_0)$ and ${\cal U}(s_3)$ are unstable. Since $H'(s)V'(s) < 0$ (resp.\ $ H'(s)V'(s) \geq 0 $), ${\cal U}(s)$ is unstable (resp.\ stable) for $ s \in (s_0,s_1) \cup (s_2,s_3) $ (resp.\ $ s \in [ s_1, s_2 ]$).

The stability of the unduloid depending on $s$ is summarized in Table~\ref{tbl:lambda2}, and values of $s_k \; (k=0,1,2,3)$ numerically obtained are presented in Table~\ref{tbl:sk}.

As mentioned before, the existence of stable unduloids for $n=7$ has not been known. Thus, the stable unduloid for $ s \in [ s_1, s_2 ]$ is found for the first time in this paper.

\subsection{Class C: $8 \leq n \leq 9$}
\label{sec:n8}

In this class the area-volume curve of an unduloid has two cusps as in Class B. The quantity $V'(s)$ has two simple zeros, but $H'(s)$ has only one simple zero. Taking into account the correspondence with Class B, we denote these zeros as follows.
\be
	V'(s_1)=V'(s_{2})=H'(s_{3})=0, \;\;\; 0<s_1<s_{2}< s_3 <1.
\label{n8s}
\ee

From the behavior of $H'(s)$, one sees that $\lambda_2(s)$ changes sign only at $s=s_3$ with criteria \eqref{cr21}--\eqref{cr23}. From this fact and the behavior of $H'(s)$ with criterion \eqref{cr1}, one can see that $\lambda_2(s) \geq 0$ (resp.\ $\lambda_2 (s) < 0$) for $s \in (0,s_3]$ (resp.\ $s \in (s_3,1)$). Therefore, ${\cal U}(s)$ is unstable for $s \in (s_3,1)$ with criterion \eqref{cr0}. Since $\lambda_2(s) \geq 0$ for $s \in (0,s_3]$, one has to see the behavior of $V'(s)$ to use criteria \eqref{cr3} and \eqref{cr4}. Since $V'(s_3) \neq 0$, ${\cal U}(s_3)$ is unstable with \eqref{cr3}. Since $H'(s)V'(s)<0$ (resp.\ $H'(s)V'(s) \geq 0$), ${\cal U}(s)$ is unstable (resp.\ stable) for $s \in (0,s_1) \cup (s_2,s_3)$ (resp.\ $ s \in [s_1, s_2] $) with \eqref{cr4}. These results are consistent with the known results in the literature, {\it i.e.}, statements (i)--(iv).

\subsection{Class D: $n \geq 10$}
\label{sec:n10}

In this class the area-volume curve of an unduloid has only one cusp. Both $H'(s)$ and $V'(s)$ have a simple zero. Taking into account the correspondence to the other classes, we denote the zeros as follows.
\be
	V'(s_2) = H'(s_3) = 0,
\;\;\;
	0<s_2<s_3<1.
\label{n10s}
\ee

From the behavior of $H'(s)$ and criteria \eqref{cr1}--\eqref{cr23}, one sees that $\lambda_2(s)$ changes sign only at $s=s_3$ and $\lambda_2(s) \geq 0$ (resp.\ $\lambda_2(s)<0$) for $s \in (0,s_3]$ (resp.\ $s \in (s_3,1)$). Therefore, ${\cal U}(s)$ for $s \in (s_3,1)$  is unstable with \eqref{cr0}. 
Since $\lambda_2(s) \geq 0$ for $s \in (0,s_3]$, we have to see the behavior of $V'(s)$ in order to use criteria \eqref{cr3} and \eqref{cr4}. Since $V'(s_3) \neq 0$, ${\cal U}(s_3)$ is unstable with \eqref{cr3}. Since $H'(s)V'(s)<0$ (resp.\ $H'(s)V'(s) \geq 0$), ${\cal U}(s)$ is unstable (resp.\ stable) for $s \in (s_2,s_3)$ (resp.\ $ s \in (0, s_2] $) with criterion \eqref{cr4}.

\section{Summary and discussions}
\label{sec:conclusion}

The equilibrium configuration of continuum supported by only surface tension ({\it i.e.}, ignoring external gravity and self-gravity) is well approximated by a CMC (constant-mean-curvature) surface that is a critical point of the variational problem extremizing the surface area while keeping the volume fixed. We have investigated the stability of CMC hypersurfaces in ${\mathbb R}^{n+2} \; (n \in {\mathbb N})$ that possibly have boundaries on two parallel hyperplanes, by examining if the CMC hypersurfaces not only extremize but also minimize the surface area amongst all nearby surfaces while keeping the volume fixed. In particular, we have focused on the stability of non-uniform liquid bridges, known as the Delaunay unduloids~\cite{Delaunay}, for which stability had been known partially~\cite{Athanassennas, Vogel, PR, Li} as statements (i)--(iv) presented in Sect.~\ref{sec:analysis}.

We have revealed the stability of unduloids for all $n \in {\mathbb N}$ and for all range of non-uniformness parameter $s \in (0,1)$, defined by Eq.~\eqref{nonuni}. After obtaining mean curvature $H$ and volume $V$ of unduloids as functions of $s$ numerically, the stability of unduloids was determined using their derivatives and stability criteria \eqref{cr0} and \eqref{cr1}--\eqref{cr4} presented in Sect.~\ref{sec:criteria}.

Although the behaviors of both $H(s)$ and $V(s)$ have played the central roles in our stability analysis, an interesting point is that the regions of $s$ where the unduloid is stable (resp.\ unstable) completely coincide with those where $V(s)$ is non-increasing (resp.\ increasing) for any $n$ (see Table~\ref{tbl:lambda2}). Therefore, the bottom line of the stability analysis is summarized without mentioning $H(s)$ as follows.

{\it
Let $s \in (0,1)$ be the non-uniformness parameter of a half period of an unduloid between two parallel hyperplanes in ${\mathbb R}^{n+2} \; (n \in {\mathbb N})$ defined by Eq.~\eqref{nonuni}. The half period of an unduloid with parameter $s$ and its bulk $(n+2)$-volume are denoted by ${\cal U}(s)$ and $V(s)$, respectively (the distance between the two hyperplanes is fixed). Then, the following (I)--(IV) hold. 
\begin{itemize}
\item[{(I)}]
For any $n \geq 1$, ${\cal U}(s)$ is stable (reps.\ unstable) if and only if $V'(s) \leq 0$ (resp.\ $V'(s) > 0$).
\item[{(II)}]
If $1 \leq n \leq 6$, then ${\cal U}(s)$ is unstable for any $s \in (0,1)$.
\item[{(III)}]
If $7 \leq n \leq 9$, there exist $ s_1$ and $s_2$ such that $\; V'(s_1)=V'(s_2)=0$ and $0<s_1 < s_2<1$. For any $s \in [s_1,s_2]$ (resp.\ $s \in (0, s_1) \cup (s_2,1)$), ${\cal U}(s)$ is stable (resp.\ unstable).
\item[{(IV)}]
If $n \ge 10$, there exists $s_2$ such that $V'(s_2)=0$ and $0<s_2<1$. For any $s \in (0,s_2]$ (resp.\ $s \in (s_2,1)$), ${\cal U}(s)$ is stable (resp.\ unstable).
\end{itemize}
}

The values of $s_1$ and $s_2$ are presented in Table~\ref{tbl:sk} with other characteristic values, $s_0$ and $s_3$ (see Eqs.~\eqref{n7s}, \eqref{n8s}, and \eqref{n10s} for the definitions).

\begin{table}
\centering
\tabcolsep=3 mm

\caption{Values of $s_k\; (k=0,1,2,3)$ for several $n$. There exists no $s_k\; (k=0,1,2,3)$ in Class A ($1 \leq n \leq  6$). The existence of $s_k \; (k=0,1,2,3)$ is the same in each class.}
\label{tbl:sk}

\vspace{5pt}

\begin{tabularx}{130mm}{c  c  c  c  c  c  c  c  c  c }
\Hline
$n$ & 7 & 8 & 9 & 10   & 11 \\
\hline
$s_0$ &  0.437 & n/a & n/a & n/a & n/a  \\
\hline
$s_1$ &  0.507 & 0.275 03 & 0.093 270 8 & n/a  & n/a \\
\hline
$s_2$ &  0.665 & 0.765 33 & 0.803 961 7 & 0.828 991 30 & 0.847 468 517 \\
\hline
$s_3$ &  0.671 & 0.765 41 & 0.803 966 2 & 0.828 991 56  & 0.847 468 533 \\
\Hline
\end{tabularx}
\end{table}

We have not paid much attention to hemispheres and cylinders since their stability structure is completely understood as mentioned in Sect.~\ref{sec:evp}. Nevertheless, let us have a look at them here, from which one can see the inevitability of the region where $V'(s) \leq 0$ for $ n \ge 8 $. As can be seen in Fig.~\ref{fig:diagrams}, the area-volume curves of hemisphere and cylinder intersect for $ n \leq 7$, but not for $n \ge 8$. A crucial reason of this is that the ratio of the volume of the critical cylinder $V(0)$ to that of the largest hemisphere ({\it i.e.}, the hemisphere which fits the interval $[z_1,z_2]$) $V(1)$, given by  
\begin{align}
	\frac{V(0)}{V(1)}
	=
	\frac{ v_{n+1} r^{n+1} L\rvert_{r=r_c} }{ \frac12 v_{n+2} L^{n+2} }
	=
\begin{cases}
	\displaystyle
	\frac{ (n+2)n^{ \frac{n+3}{2} } (n-1)! }{ 2^{n-1}\pi^{n+1}(n+1)\left[ \left( \frac{n-1}{2} \right)! \right]^2 }
&
	(n:{\rm odd})
\\
	\displaystyle
	\frac{ 2^{n+1}(n+2)n^{ \frac{n-1}{2} } \left[ \left( \frac{n}{2} \right)! \right]^2 }{ \pi^{n+2} (n+1) (n-1)! }
&
	(n:{\rm even})
\end{cases},
\end{align}
increases with $n$ and becomes larger than unity for $n \geq 8$ (see Table~\ref{tbl:VMSC}). Namely, for $n \geq 8$ the branch of unduloids emanating from the critical cylinder at $s=0$ must have a region where the volume decreases to reach the largest hemisphere at $s=1$. Although dimension $n=7$, at and above which the stable unduloid exists, differs from this critical number of dimension $n=8$ by one, their closeness is clearly not a coincidence.

\begin{table}
\centering
\tabcolsep=2 mm

\caption{The ratio of the volume of critical cylinder $V(0)$ to the volume  of largest hemisphere $V(1)$, that fits the interval $[z_1,z_2]$. This quantity exceeds the unity at $n=8$.}
\label{tbl:VMSC}

\vspace{5pt}

\begin{tabularx}{125mm}{c  c  c  c  c  c  c  c  c  c }
\Hline
$n$ & 1 & $\cdots$ & 6 & 7 & 8 & 9 & 10 & 11 \\
\hline
$V(0)/V(1)$ & 0.152 & $\cdots$ & 0.408 & 0.623 & 1.005 & 1.707 & 3.033 & 5.621 \\
\Hline
\end{tabularx}
\end{table}

In passing, let us point out that the area-volume curve deforms continuously if $n$ changes continuously. As $n$ increases from 1, the `swallowtail' (two cusps) of the area-volume curve appears at $n \simeq 7$. As $n$ increases further, the swallowtail becomes large. In other words, $s_1$ decreases to approach $0$ and $s_2$ increases to approach $1$. Indeed, $s_1$ decreases as $n$ increases to vanish finally at $n \simeq 10$. As far as we know, $s_2$ continues to increase but does not vanish for arbitrarily large $n \gg 1$, which is consistent with statement (i). If one treats $n$ as a continuous parameter and examines the stability for non-integer $n$, which seems to bring no technical problem, the behaviors of area-volume curve and stability structure expected and described above would be observed.

In this paper, the stability of unduloids was determined by the behaviors of $H(s)$ and $V(s)$, which were obtained by numerical integration. Therefore, the correctness of the conclusions is based on that of these numerical computations. It is noted that one needs highly accurate computation to show that $ s_2 < s_3 $ holds ($s_2$ and $s_3$ are defined as the zeros of $V'(s)$ and $H'(s)$, respectively) for $n \geq 8$. For example, $s_3/s_2 - 1 \simeq + 5.6 \times 10^{-6} $ for $n=9$ by our computation, and this quantity seems to decrease further as $n$ increases. Nevertheless, we assumed that $s_2 < s_3$ continues to hold for arbitrarily large $n$, otherwise our conclusions on the stability might be different from those presented in the text. Therefore, any analytic method or alternative numerical methods that guarantee accuracy will be helpful to confirm the results in this paper.

Related to the results of this paper, one of the most interesting problems would be to investigate the implications to dynamical problems. While this was partially worked by one of the present authors in~\cite{Miyamoto:2008uf} using the surface-diffusion equation~\cite{Mullins, BBW}, there are still many things to do in this direction.

We remark that the stability of black strings qualitatively exhibits a similar dependence on the dimension. Suppose a $D$-dimensional vacuum spacetime ($D\geq 5$) with one spacelike dimension compactified to a circle $S^1$. Then, there exist non-uniform black strings of which horizon topology is $S^{D-3} \times S^1$. The stability of such black strings has been examined using the thermodynamic criterion, and argued as follows~\cite{Figueras:2012xj}. If $ 5 \leq D \leq 11$, all non-uniform black strings are unstable. If $12 \leq D \leq 13$, there exists a critical non-uniformness below (resp.\ above) which the non-uniform black strings are unstable (resp.\ stable). If $D \geq 14$, all black strings are stable. We are not so surprised at the similarity of stability between these black objects and CMC hypersurfaces since it was shown that the event horizon of a black hole is approximated by a CMC hypersurface in the large-dimension limit~\cite{Emparan:2015hwa}. Nevertheless, it is still interesting to pursue the similarity from various points of view such as the fluid/gravity correspondence~\cite{Aharony:2005bm, Bhattacharyya:2008jc} and the gauge/gravity correspondence~\cite{Azuma:2014cfa}.

\subsection*{Acknowledgments}

The authors wish to thank the anonymous reviewer for many useful comments on the earlier version of this paper. 
This work was partially supported by JST CREST GrantNumber JPMJCR1911. 
It was supported in part also by JSPS KAKENHI Grant Numbers 
JP18H04487, JP20H01801, JP20H04642 (K.M.),
JP18K03652,
and
JP22K03623 (U.M.).

\appendix

\section{Mathematical propositions and their proofs}
\label{sec:math}

\subsection{Stability of a half period and one period of an unduloid}
\label{period}

Here, we prove
\begin{proposition}\label{peri}
If a half period of an unduloid is stable (resp.\ unstable) between two parallel hyperplanes in ${\mathbb R}^{n+2} \; (n \in {\mathbb N})$, one period of the unduloid is stable (resp.\ unstable) in ${\mathbb R}^{n+1} \times S^1$, and {\it vice versa}. 
\end{proposition}

\noindent {\it Proof.} \ 
Let $X$ be a half period of an unduloid $\mathcal U$ with the $z$-axis as its axis of revolution which is generated by a curve
\begin{equation}\label{gen1}
(z, h(z)), \ h(z)>0, \quad 0\le z \le z_2.
\end{equation}
This implies that $X$  is perpendicular to the hyperplanes $\Pi_0=\{z=0\}$, $\Pi_2=\{z=z_2\}$. 
Without loss of generality, we may assume that $z=0$, $z=z_2$  corresponds to the bulge and the neck of $\mathcal U$, respectively. 
Denote by $Y$ the one period of $\mathcal U$ that is generated by 
\begin{equation}\label{gen12}
(z, h(z)), \ h(z)>0, \quad -z_2\le z \le z_2.
\end{equation}

Assume that $X$ is unstable. 
Then, there exists a volume-preserving variation $X(\epsilon)$ of $X$ such that 
$$
\frac{d^2 A(X(\epsilon))}{d\epsilon^2}\Big\vert_{\epsilon=0}<0
$$
holds. By reflection with respect to $\Pi_0$, we get a volume-preserving variation $Y(\epsilon)$ of $Y$ which satisfies 
$$
\frac{d^2 A(Y(\epsilon))}{d\epsilon^2}\Big\vert_{\epsilon=0}
=2\frac{d^2 A(X(\epsilon))}{d\epsilon^2}\Big\vert_{\epsilon=0}<0. 
$$
This implies that $Y$ is also unstable. 

Assume now that $Y$ is unstable. 
Then, there exists a volume-preserving variation $Y(\epsilon)$ of $Y$ such that 
$$
\frac{d^2 A(Y(\epsilon))}{d\epsilon^2}\Big\vert_{\epsilon=0}<0
$$
holds. 
Let $\hat{Y}(\epsilon)$ be the Steiner symmetrization of $Y(\epsilon)$ with respect to $\Pi_0$, that is, $\hat{Y}(\epsilon)$ is a hypersurface defined by the conditions (i) and (ii) below. 
Note that we consider only hypersurfaces close to $Y$. 
Set $\Pi_{-2}=\{z=-z_2\}$. 
Denote by $G(\epsilon)$, $\hat{G}(\epsilon)$ the closed domains bounded by $Y(\epsilon)\cup \Pi_{-2}\cup \Pi_2$, $\hat{Y}(\epsilon)\cup \Pi_{-2}\cup \Pi_2$, respectively. For each point $P\in \Pi_0$, denote by $L_P$ the straight line that passes through $P$ and is perpendicular to $\Pi_0$. Define two straight line segments by $\hat{\Gamma}(\epsilon):=L_P\cap \hat{G}(\epsilon)$, $\Gamma(\epsilon):=L_P\cap G(\epsilon)$. 
Note the following. 

(i) The lengths of $\hat{\Gamma}(\epsilon)$ and $\Gamma(\epsilon)$ are the same, ($\forall P, \epsilon$).

(ii) The middle point of $\hat{\Gamma}(\epsilon)$ lies on $\Pi_0$, ($\forall\epsilon$). 

\noindent Then, it is well-known that 

(a) $V(G(\epsilon))=V(\hat{G}(\epsilon))$ holds, ($\forall\epsilon$), 

(b) $A(Y(\epsilon))\ge A(\hat{Y}(\epsilon))$ holds, ($\forall\epsilon$)

\noindent hold ({\it cf.}~\cite[Note A]{PS}).
Therefore, $\hat{Y}(\epsilon)$ is a volume-preserving variation of $Y$ such that it is symmetric with respect to $\Pi_0$ and such that 
$$
\frac{d^2 A(\hat{Y}(\epsilon))}{d\epsilon^2}\Big\vert_{\epsilon=0}
\le
\frac{d^2 A(Y(\epsilon))}{d\epsilon^2}\Big\vert_{\epsilon=0}
<0
$$
holds. 
The restriction $\hat{X}(\epsilon)$ of $\hat{Y}(\epsilon)$ to $\{0 \le z \le z_2\}$ is a volume-preserving variation of $X$ such that 
$$
\frac{d^2 A(\hat{X}(\epsilon))}{d\epsilon^2}\Big\vert_{\epsilon=0}
=
\frac{1}{2}
\frac{d^2 A(\hat{Y}(\epsilon))}{d\epsilon^2}\Big\vert_{\epsilon=0}<0
$$
holds. 
Hence $X$ is unstable. 
\hfill$\Box$

\subsection{Stability criteria for axially symmetric equilibrium hypersurfaces}
\label{sst}

Let 
\begin{equation}\label{gen}
(z, h(z)), \ h(z)>0, \quad z_1\le z \le z_2
\end{equation}
define an axially-symmetric equilibrium hypersurface $X$ with the $z$-axis as its axis of revolution, 
that is, $X$ is a part of either a cylinder or an unduloid with generating curve (\ref{gen}) and it is perpendicular to the hyperplanes $\Pi_i=\{z=z_i\}$, ($i=1, 2$).  Then, one can show the following lemma.

\begin{lemma}\label{lsym2}
$X$ is stable if and only if $X$ is stable for axially-symmetric variations.
\end{lemma}

\noindent {\it Proof}. 
Assume that $X$ is unstable. 
Then, there exists a volume-preserving variation $X(\epsilon)$ of $X$ such that 
$$
\frac{d^2 A(X(\epsilon))}{d\epsilon^2}\Big\vert_{\epsilon=0}<0
$$
holds.
Let $\hat{X}(\epsilon)$ be the Schwarz symmetrization of $X(\epsilon)$, that is $\hat{X}(\epsilon)$ is an axially-symmetric hypersurface defined by the conditions (i) and (ii) below. 
Note that we may assume that $X(\epsilon)$ does not have self-intersection and it is contained in the closed domain bounded by $\Pi_1$, $\Pi_2$, because we consider only hypersurfaces close to $X$. 
Denote by $G(\epsilon)$, $\hat{G}(\epsilon)$ the closed domains bounded by $X(\epsilon)\cup \Pi_1\cup \Pi_2$, $\hat{X}(\epsilon)\cup \Pi_1\cup \Pi_2$, respectively. For each hyperplane $\Pi_c:=\{z=c\}$, ($z_1\le c \le z_2$), set $\hat{D}(\epsilon):=\Pi_c\cap \hat{G}(\epsilon)$, then it is a round $(n+1)$-ball. 

(i) $\hat{D}(\epsilon)$ has the same $(n+1)$-volume as $D(\epsilon):=\Pi_c\cap G(\epsilon)$, ($\forall c, \epsilon$).

(ii) The center of $\hat{D}(\epsilon)$ lies on the $z$-axis. 

\noindent Then, it is well-known that 

(a) $V(G(\epsilon))=V(\hat{G}(\epsilon))$ holds, ($\forall\epsilon$), 

(b) $A(X(\epsilon))\ge A(\hat{X}(\epsilon))$ holds, ($\forall\epsilon$)

\noindent hold ({\it cf.}~\cite[Note A]{PS}).
Therefore, $\hat{X}(\epsilon)$ is a volume-preserving axially-symmetric variation of $X$ such that 
$$
\frac{d^2 A(\hat{X}(\epsilon))}{d\epsilon^2}\Big\vert_{\epsilon=0}<0
$$
holds.
Hence $X$ is unstable for axially-symmetric variations. 

The opposite direction is trivial. 
\hfill $\Box$

Now we give two criteria for the stability of axially-symmetric equilibrium hypersurfaces. 
The first criterion (Lemma \ref{slem1}) will be proved by using the second criterion (Lemma \ref{stability1}) at the end of this subsection.

\begin{lemma}[First stability criterion] \label{slem1} 
Assume that $X(s)$ is a one-parameter smooth family of axially-symmetric equilibrium hypersurfaces generated by the curves
\begin{equation}\label{gen3}
(z, h(z, s)), \ h(z, s)>0, \quad z_1\le z \le z_2,
\end{equation} 
that is, $X(s)$ are half periods of unduloids, and we assume that $s$ is the parameter defined by (\ref{nonuni}). Denote by $H(s)$, $V(s)$ the mean curvature, the enclosed $(n+2)$-dimensional volume of $X(s)$, respectively. 
Denote by $\lambda_i(s)$ the $i$-th eigenvalue of the eigenvalue problem (\ref{evp}) for $X(s)$.

\noindent {\rm (I)} \ If $\lambda_1(s) \ge 0$, 
then $X(s)$ is stable. 

\noindent {\rm (II)} If $\lambda_1(s) < 0 < \lambda_2(s)$, then the following {\rm (II-1)'} and {\rm (II-2)'} hold. 

{\rm (II-1)'} If $H'(s)V'(s)\ge 0$, then $X(s)$ is stable.

{\rm (II-2)'} If $H'(s)V'(s)< 0$, then $X(s)$ is unstable.

\noindent {\rm (III)} If $\lambda_2(s)=0$, then the following {\rm (III-a)} and {\rm (III-b)} hold{\rm :}

{\rm (III-a)} Assume that $H'(s)= 0$ holds. If $V'(s)\ne 0$, then $X(s)$ is unstable.

{\rm (III-b)} Assume that $H'(s)\ne 0$ holds.

\quad {\rm (III-b1)} If $H'(s)V'(s)\ge 0$, then $X(s)$ is stable.

\quad {\rm (III-b2)} If $H'(s)V'(s)< 0$, then $X(s)$ is unstable.

\noindent {\rm (IV)} If $\lambda_2(s) < 0$, 
then $X(s)$ is unstable.
\end{lemma}

\begin{remark}\label{remv} 
In (III-a) in the above theorem, we assumed $H'(s)=0$ and $V'(s)\ne0$. If $\lambda_2(s)=0$, $H'(s)=0$ and $V'(s)=0$, then all of the assumptions in (III-B) of Lemma \ref{stability1} are satisfied (see the proof of Lemma \ref{slem1}). And hence, there exists such a function $u$ indicated there and (III-B1), (III-B2) hold.
\end{remark}

In view of Lemma \ref{lsym2}, 
the following lemma is proved by a modification of the proof of Theorem 1.3 in \cite{Ko2}.
\begin{lemma}[Second stability criterion] 
\label{stability1} 
Let $X$ be an axially-symmetric equilibrium hypersurface generated by the curve
\begin{equation}\label{gen2}
(z, h(z)), \ h(z)>0, \quad z_1\le z \le z_2.
\end{equation}
\begin{itemize}
\item[{\rm (I)}] \ If $\lambda_1 \ge 0$, then $X$ is stable.
\smallskip

\item[{\rm (II)}] If $\lambda_1 < 0 < \lambda_2$,
then there exists a uniquely determined $C^\infty$ 
function $u :[z_1, z_2] \to {\mathbb R}$ which
satisfies ${\cal L}u = h^n$ and $u'(z_1)=u'(z_2)=0$, and the following statements hold.\smallskip

\begin{itemize}
\item[{\rm (II-1)}] If $ \int_{z_1}^{z_2} uh^n \;dz \ge 0$, then $X$ is stable.\smallskip

\item[{\rm (II-2)}] If $ \int_{z_1}^{z_2} uh^n \;dz < 0$, then $X$ is unstable.
\end{itemize}\medskip

\item[{\rm (III)}] If $\lambda_1 < 0 = \lambda_2$,
then the following statements hold:\smallskip

\begin{itemize}
\item[{\rm (III-A)}] If there exists a $\lambda_2$-eigenfunction $e$
which satisfies $ \int_{z_1}^{z_2} eh^n \;dz \ne 0$, then $X$ is unstable.
\smallskip
\item[{\rm (III-B)}] If $ \int_{z_1}^{z_2} eh^n \;dz = 0$ for any
$\lambda_2$-eigenfunction $e$, then there exists a uniquely determined $C^\infty$ 
function $u :[z_1, z_2] \to {\mathbb R}$ which
satisfies ${\cal L}u = h^n$,  $u'(z_1)=u'(z_2)=0$, and 
$ \int_{z_1}^{z_2} euh^n \;dz = 0$ holds for any
$\lambda_2$-eigenfunction $e$. And the following statements hold:\smallskip

\begin{itemize}
\item[{\rm (III-B1)}] If $ \int_{z_1}^{z_2} uh^n \;dz \ge 0$, then $X$ is stable.\smallskip

\item[{\rm (III-B2)}] If $ \int_{z_1}^{z_2} uh^n \;dz < 0$, then $X$ is unstable.
\end{itemize}
\end{itemize}\medskip

\item[{\rm (IV)}] If $\lambda_2 < 0$,  then $X$ is unstable.
\end{itemize}
\end{lemma}

The following observation will be used to prove Lemma \ref{slem1}. 

\begin{lemma}\label{hs} 
If $X(s)$ defined by (\ref{gen3}) are half periods of unduloids and $s$ is the parameter defined by (\ref{nonuni}), then
for any fixed $s$, 
\begin{equation}\label{eq-hs}
h_s(z, s)\not\equiv 0
\end{equation}
holds for all $z \in [z_1, z_2]$, where $\displaystyle 
h_s=\partial h/\partial s$.
\end{lemma}

\noindent{\it Proof}. \ 
Set
\be
	\rho := 1-s.
\label{nonuni3}
\ee
It is sufficient to prove that 
for any fixed $\rho$, 
\begin{equation}\label{eq-hr}
h_\rho(z, \rho)\not\equiv 0
\end{equation}
holds for all $z \in [z_1, z_2]$, 

Now we may assume that 
\be
\rho=\frac{h(z_1, \rho)}{h(z_2, \rho)}
\label{eq-r}
\ee
holds. 
Differentiating both sides of (\ref{eq-r}) with respect to $\rho$, we obtain
\be
1=\frac{h_\rho(z_1, \rho)h(z_2, \rho)-h(z_1, \rho)h_\rho(z_2, \rho)}{(h(z_2, \rho))^2}.
\label{dr}
\ee
This implies that
\begin{equation}\label{eq-hr2}
h_\rho(z, \rho)= 0, \quad \forall z \in [z_1, z_2]
\end{equation}
never occurs. 
\hfill $\Box$

\noindent{\it Proof of Lemma \ref{slem1}}. \ 
(I) and (IV) are the same as those in Lemma \ref{stability1}. So we will prove (II) and (III). As Eq.\ (\ref{ell}), one can show that 
\be
	{\cal L} h_s 
	=(n+1)h^nH'(s).
	\label{ell2}
\ee
Note that $H'(s)$ depends only on $s$. 

First, we prove (II). Since $h_s\not\equiv 0$ (Lemma \ref{hs}), 
and since zero is not an eigenvalue of (\ref{evp}), (\ref{ell2}) implies $H'(s)\ne 0$. Hence, the function $u$ given in (II) of Lemma \ref{stability1} satisfies
\begin{equation}\label{u1}
u=\frac{h_s}{(n+1)H'(s)}.
\end{equation}
From Eq.~\eqref{volume}, the following holds (changing the variable from $\epsilon$ to $s$),
\begin{equation}\label{V12}
	V'(s)
	=
	a_n \int_{z_1}^{z_2} h^n h_s dz.
\end{equation}
Equation (\ref{u1}) with (\ref{V12}) gives
\begin{equation}\label{V13}
\int_{z_1}^{z_2} u h^n dz=\frac{V'(s)}{(n+1) a_n H'(s)},
\end{equation}
which shows that (II) in Lemma \ref{slem1} is equivalent to (II) in Lemma \ref{stability1}. 

Next, we prove (III). 
Let $e$ be an eigenfunction belonging to $\lambda_2=0$. Then, 
\begin{equation}\label{self}
\int_{z_1}^{z_2} e {\cal\; L}h_s dz=\int_{z_1}^{z_2} h_s{\cal L}e  \;dz=0. 
\end{equation}
On the other hand, using (\ref{ell2}), we have
\begin{equation}\label{self2}
\int_{z_1}^{z_2} e {\cal\; L}h_s dz=(n+1)H'(s)\int_{z_1}^{z_2} eh^n  \;dz. 
\end{equation}
Equation (\ref{self}) combined with Eq.\ (\ref{self2}) gives
\begin{equation}\label{self3}
H'(s)\int_{z_1}^{z_2} eh^n  \;dz=0. 
\end{equation}

First, assume that $H'(s)=0$ holds. 
Then, since $h_s\not\equiv 0$ 
(Lemma \ref{hs}), Eq.\ (\ref{ell2}) implies that $h_s$ is an eigenfunction belonging to $\lambda_2=0$. 
Since each eigenspace is one-dimensional, 
\begin{equation}\label{ef1}
h_s=ce, \quad \exists c \in {\mathbb R}\setminus \{0\}.
\end{equation}
Hence, 
\begin{equation}\label{V'}
V'(s)=a_n\int_{z_1}^{z_2}h_s h^n \;dz=c a_n\int_{z_1}^{z_2}eh^n\;dz.
\end{equation}
Therefore, $V'(s)\ne 0$ is equivalent to $ \int_{z_1}^{z_2}eh^n\;dz\ne 0$. From (III-A) in Lemma \ref{stability1}, $X(s)$ is unstable. This gives (III-a). 

Lastly, we prove (III-b). 
Assume that $H'(s)\ne 0$ holds. From (\ref{self3}), there holds
\begin{equation}\label{self4}
\int_{z_1}^{z_2} eh^n  \;dz=0. 
\end{equation}
Take the function $u$ that is uniquely defined in (III-B) of Lemma \ref{stability1}. 
Using (\ref{ell2}), $h_s$ is written as 
\begin{equation}\label{dec}
h_s=(n+1)H'(s)u + be, \quad \exists b \in {\mathbb R}.
\end{equation}
Then, 
\begin{equation}\label{V'2}
V'(s)=a_n\int_{z_1}^{z_2}h_s h^n \;dz=(n+1)a_n H'(s)\int_{z_1}^{z_2}uh^n\;dz
\end{equation}
holds, which gives
\begin{equation}\label{V'3}
\int_{z_1}^{z_2}uh^n\;dz=\frac{V'(s)}{(n+1)a_n H'(s)}.
\end{equation}
Hence, from (III-B) of Lemma \ref{stability1}, we obtain (III-b).  
\hfill $\Box$

\subsection{Negativity (resp.\ positivity) of the first (resp.\ third) eigenvalue for unduloids}
\label{sec:lambda13}
 
In addition to the eigenvalue problem \eqref{evp}, 
we consider also the following eigenvalue problem with Dirichlet boundary condition:
\begin{align}
\begin{split}
	{\cal L} \psi_i (z) &= -\lambda_i \psi_i (z), \quad z_1\le z \le z_2,
\\
	\psi_{i}(z_1) & = \psi_{i}(z_2) = 0,
\end{split}
\label{evp3}
\end{align}
and denote its $i$-th eigenvalue by $\lambda_i^0[z_1, z_2]=\lambda_i^0$. 
Then, it is well-known that $\lambda_1^0<\lambda_2^0 < \cdots$ and 
\be
\lambda_i<\lambda_i^0, \quad \forall i \in {\mathbb N}
\label{ineq}
\ee
hold. 
Also recall that, 
since ${\cal L}$ is a Sturm-Liouville operator, each of $\varphi_i (z)$ and $\psi_i (z)$ has exactly $i-1$ zeros in $( z_1,z_2 )$.
Recall that Eq.\ (\ref{ell}) holds and consider the parallel translation
$$
	{\cal U}(\epsilon):(z+\epsilon, h(z)), \quad z_1\le z \le z_2
$$
of the unduloid
$$
	{\cal U}:(z, h(z)), \quad z_1\le z \le z_2.
$$ 
Then, the mean curvature $H(\epsilon)$ is the same as the mean curvature $H_0$ of ${\cal U}$. ${\cal U}(\epsilon)$ can be represented as
$$
{\cal U}(\epsilon): (z, h(z-\epsilon)), \quad z_1+\epsilon\le z \le z_2+\epsilon.
$$
Hence, we have from (\ref{ell}) that
\be
0={\cal L} h_1={\cal L}(-h_z)=-{\cal L}(h_z).
\label{efun}
\ee
Since ${\cal U}$ is a half period of an unduloid, 
$$
h_z(z_1)=h_z(z_2)=0
$$
holds and we may assume that $h_z>0$ on $z_1<z<z_2$. Hence, $h_z$ is an eigenfunction of (\ref{evp3}) and the corresponding eigenvalue zero is the first eigenvalue $\lambda_1^0$. Hence, by (\ref{ineq}), 
\be
\lambda_1<\lambda_1^0=0
\label{ineq2}
\ee
holds. 
Next, we assume that
\be
\lambda_3\le 0
\label{ineq3}
\ee
holds. Then, eigenfunction $\varphi_3 (z)$ has exactly two zeros $\zeta_1, \zeta_2$, ($\zeta_1<\zeta_2$), in $(z_1,z_2)$. Hence, by the monotonicity of the eigenvalues of the problem (\ref{evp3}) with respect to the domain, we have
$$0=\lambda_1^0[z_1, z_2]<\lambda_1^0[\zeta_1, \zeta_2]=0, 
$$
which is a contradiction. Hence, $\lambda_3$ must be positive.

\subsection{Existence of bifurcation and estimate of eigenvalues in a bifurcation branch}
\label{sbif}

Assume that 
$I\subset {\mathbb R}$ 
is an non-empty open interval, and 
$$
(z, h(z, \epsilon)), \ h(z, \epsilon)>0, \quad z_1\le z \le z_2, \ \epsilon \in I \subset {\mathbb R}
$$
defines a smooth one-parameter family of axially-symmetric equilibrium hypersurfaces $\{X(\epsilon)\}_{\epsilon \in I}$, 
that is, each $X(\epsilon)$ is a part of either a cylinder or an unduloid and it is perpendicular to the hyperplanes $\Pi_i=\{z=z_i\}$, ($i=1, 2$), and $X(\epsilon)$ is of $C^\infty$ in $\epsilon$.
Denote by $H(\epsilon)$ the mean curvature of $X(\epsilon)$. 
Denote by $\lambda_i(X(\epsilon))$ 
the $i$-th eigenvalue of the problem 
(\ref{evp}) for $X(\epsilon)$. 

Now we define the concept ``bifurcation instant''.

\begin{definition}\label{bifdef}
For $\overline{\epsilon}\in I$, we say that $\overline{\epsilon}$ is a \emph{bifurcation
instant} for the family $\{X(\epsilon)\}_{\epsilon \in I}$ if there exists a sequence $\{\epsilon_k\}_{k\in{\mathbb N}}$ in $I$ and a sequence
$\{Y_k\}_{k\in {\mathbb N}}$ such that:
\begin{itemize}
\item[(i)] $\epsilon_k\to\overline \epsilon$ as $k\to\infty$.
\item[(ii)] Each $Y_k$ is an axially-symmetric equilibrium hypersurface that is defined by 
$$
(z, h_k(z)), \ h_k(z)>0, \quad z_1\le z \le z_2,
$$
and the mean curvature of $Y_k$ is equal to $H({\epsilon_k})$ for all $k$.
\item[(iii)] $h_k(z)\to h(z, \overline\epsilon)$, ($z_1\le z \le z_2$), as $k\to\infty$.
\item[(iv)] $h_k(\ast)\ne h(\ast, \epsilon)$ for all $\epsilon \in I$ and $k\in{\mathbb N}$.
\end{itemize}
In other words, $\overline \epsilon$ is a bifurcation instant for the family $\{X(\epsilon)\}_{\epsilon\in I}$ if $X({\overline \epsilon})$ is an accumulation
of equilibrium hypersurfaces  that are not congruent to any of the hypersurface
of the family $\{X(\epsilon)\}_{\epsilon\in I}$.
\end{definition}

\begin{remark}\label{rbif}{\rm
The following Theorems \ref{bifurcation1}, \ref{bifurcation-stabilityII} are proved by modifications of the proofs of Theorems 1.1, 6.4 in \cite{KPP2017}, respectively.

}\end{remark}

\begin{theorem}[Existence of bifurcation]\label{bifurcation1}
For simplicity, we assume that $I=(-\epsilon_0, \epsilon_0)\subset {\mathbb R}$ holds. 
Assume
\begin{itemize}
\item[(i)]   $H'(0)\ne 0$.
\item[(ii)] $\lambda_i(X(0))=0$ for some $i \in {\mathbb N}$, and $e$ is an eigenfunction belonging to zero eigenvalue.
\end{itemize}
Then, $\int_\Sigma e\:\mathrm d\Sigma=0$,
and there exists a differentiable map $(-\epsilon_1, \epsilon_1)\ni \epsilon\mapsto\lambda(\epsilon)\in\mathbb R$,  with $0<\epsilon_1\le\epsilon_0$,
such that $\lambda(0)=0$, $\lambda(\epsilon)$ is a simple eigenvalue of the eigenvalue  problem 
(\ref{evp}) for $X(\epsilon)$, 
and there is no other eigenvalue of (\ref{evp}) near $0$.

Assume further that $\lambda'(0)\ne0$ holds. 
Then there is a unique smooth bifurcation branch $\{Y(t)\}_t$ of
axially-symmetric equilibrium hypersurfaces issuing at $X(0)$. More precisely,
let $E^\perp$ be the orthogonal complement of $E:=\{ae \; \vert \;a\in{\mathbb R}\}$ in $C^{\infty}([z_1, z_2])$ with respect to the $L^2$ inner product.
Then, there exist an open interval $\hat I\subset\mathbb R$ with $0 \in {\hat I}$,  and $C^1$ functions $\zeta:{\hat I} \to E^\perp$ and $\epsilon:{\hat I}\to {\mathbb R}$, such that $\epsilon(0)=0$, $\zeta(0)=0$, and $Y(t)$ is given by ${\hat h}(z, t):=h(z, \epsilon(t))+t e(z)+t\zeta(t)(z)$ with mean curvature ${\hat H}(t):=H\big(\epsilon(t)\big)$.

Moreover, the hypersurfaces $\{X(\epsilon) : \epsilon \in I\}$ and $\{Y(t) : t \in {\hat I}\}$ are pairwise distinct except for $X(0)=Y(0)$.
\end{theorem}

\begin{theorem}\label{bifurcation-stabilityII}
Under the assumptions of Theorem~\ref{bifurcation1},
denote by $\{Y(s)\}_{s\in\hat I}$ the bifurcating branch of axially-symmetric equilibrium hypersurfaces given in Theorem~\ref{bifurcation1}.
Let $\hat{H}(s)$ be the mean curvature of $Y(s)$, and $\mu(s)$ the eigenvalue for the Jacobi operator ${\cal L}_{Y(s)}$ which was defined by (\ref{jacobi}). 
We may assume that $H'(0)>0$ holds, by changing the parameter $t$ to $-t$ if necessary.

Then, the following statements are true.
\begin{itemize}
\item[(i)] If $\hat H'(s)=0$ for $s$ near $0$ ({\it i.e.}, if $\hat H$ is locally constant),
then $\mu(s)=0$ for s near $0$;
\item[(ii)] If $\hat H'(s)\ne0$ for $s>0$ small, then, for a sufficiently small
$s_0>0$, on each interval $[-s_0,0)$ and $(0,s_0]$, $\mu(s)>0$ if $\lambda'(0) s \hat H'(s)<0$, and $\mu(s)<0$
if $\lambda'(0) s \hat H'(s)>0$.
In particular, supercritical and subcritical pitchfork bifurcations correspond to the cases
where $s \hat H'(s)$ does not change sign at $s=0$ ({\it cf.}~Fig. \ref{fig:bif}), and transcritical
bifurcation occurs when $s\hat H'(s)$ changes sign at $s=0$.
\end{itemize}
\end{theorem}

\subsection{Correspondence of sign change between the eigenvalue and mean-curvature derivative}
\label{sign} 

In this section we prove (\ref{cr21}) and (\ref{cr22}). 

Assume that $\{X(\epsilon)\}_{\epsilon \in I}$ satisfies the same assumptions as that in Section \ref{sbif}. Denote by $H(\epsilon)$ the mean curvature of $X(\epsilon)$, and by $V(\epsilon)$ the enclosed $(n+2)$-dimensional volume by $X(\epsilon)$. Denote by $\lambda_i(\epsilon)$ the $i$-th eigenvalue of the problem 
(\ref{evp}) for $X(\epsilon)$.

The criterion (\ref{cr21}) is proved as follows.

\noindent {\it Proof of the criterion (\ref{cr21}).} \
 Assume that $X(s_0)$ is a part of an unduloid. 
Assume also that 
$H' \neq 0$ holds at $s_0$. 
If $\lambda_2$ changes sign at $s_0$, we can see that $s_0$ is a bifurcation instant in the same way as in the proof of Proposition 2.14 in \cite{KPP2015} which is 
an application of \cite[Theorem~2.1]{SW1990}. 
However, in our variational problem, there is no bifurcation from any unduloid $X(\epsilon)$, which is a contradiction. 
Therefore, $\lambda_2$ does not change sign at $s_0$, which proves the criterion (\ref{cr21}).
\hfill $\Box$

The criterion (\ref{cr22}) is given by the following Lemma \ref{mad}.

\begin{lemma}\label{mad}
Assume that, for a fixed $\epsilon$,  $H'(\epsilon)=0$, $H''(\epsilon)\ne 0$, $V'(\epsilon)\ne 0$, and $\lambda_j(\epsilon)=0$ for some $j \in {\mathbb N}$. Then, 
there exists a non-zero real number $\alpha$ such that
\begin{equation}\label{key}
\alpha^2\lambda_j'(\epsilon)=-(n+1)H''(\epsilon)V'(\epsilon)
\end{equation}
holds. In particular, $\lambda_j$ changes sign at $\epsilon$. 
\end{lemma}

\noindent{\it Proof}. \ 
The formula (2.8) in \cite{Maddocks} is about a functional $F: {\cal H} \times {\mathbb R} \to {\mathbb R}$, where ${\cal H}$ is a real Hilbert space. In our case, let
${\cal H}$ be the space of real-valued $C^\infty$ functions on the interval $[z_1, z_2] \subset {\mathbb R}$ with inner product
$$
\langle f, g\rangle:=\int_{z_1}^{z_2} f(z)g(z)h(z, \epsilon)^n\;dz.
$$
For any axially-symmetric (not necessarily equilibrium)  hypersurface $X_h$ generated by 
\begin{equation}\label{gen4}
(z, h(z)), \ h(z)>0, \quad z_1\le z \le z_2,
\end{equation}
set
$$
F(h, a) := A(X_h)-aV(X_h),
$$
and we define another parameter $H$ as
$$
-a:=(n+1)H.
$$
Then, the equation (\ref{key}) is equivalent to the formula (2.8) in \cite{Maddocks}. 
\hfill $\Box$

\subsection{Equivalence between the vanishing of the eigenvalue and mean-curvature derivative}
\label{vanishment} 

\begin{proposition}\label{l2}
Let $X(s)$ be a one-parameter smooth family of half periods of unduloids with mean curvature $H(s)$ generated by the curves
\begin{equation}\label{gen5}
(z, h(z, s)), \ h(z, s)>0, \quad 0\le z \le z_2, \quad s \in (s_0-\delta, s_0+\delta), \ \delta>0
\end{equation} 
with parameter $s$ defined by (\ref{nonuni}). Then $H'(s_0) = 0$ if and only if $\lambda_2(s_0)=0$ holds.
\end{proposition}

\noindent {\it Proof.} \ 
We prove the following (i), (ii) one by one. 

(i) $H'(s_0) = 0$  $\Rightarrow$ $\lambda_2(s_0)=0$

(ii) $\lambda_2(s_0)=0$ $\Rightarrow$ $H'(s_0) = 0$
\newline 
First, we prove (i).  From Eq.\ (\ref{ell2}) and Lemma \ref{hs}, if $H'(s_0) = 0$, then $0$ is an eigenvalue. This with (\ref{lambda13}) implies that $\lambda_2(s_0)=0$ holds.

Next, we prove (ii). Let the mean curvature of $X(s_0)$ be $H_0$. Then, we have a one-parameter smooth family $\{\hat{X}(s)\}$ with $\hat{X}(s_0)=X(s_0)$ of half period of unduloids with mean curvature $H_0$ generated by the curves
\begin{equation}\label{gen6}
(z, \hat{h}(z, s)), \ \hat{h}(z, s)>0, \quad 0\le z \le \zeta(s), \quad s \in (s_0-\delta', s_0+\delta'), \ \delta'>0.
\end{equation}
Note $\hat{h}(\ast, s_0)=h(\ast, s_0)$, and we denote it by $h_0$.

Now, for a half period of an unduloid generated by the curve
\begin{equation}\label{gen7}
(z, h(z)), \ h(z)>0, \quad 0\le z \le \zeta, 
\end{equation}
denote by $H[h]$ its mean curvature. Consider the equation
\begin{equation}\label{eqn}
H[\hat{h}(\ast, s)]=H_0.
\end{equation}
Then we have, using Eq.\ (\ref{ell2}), 
\begin{equation}\label{mcd}
{\cal L} \hat{h}_s\Big\vert_{s=s_0}=h_0^n(n+1)\frac{\partial H[\hat{h}(\ast, s)]}{\partial s}\Big\vert_{s=s_0}=0.
\end{equation}
Since $\hat{X}(s)$ is a suitable homothety of $X(s)$, there exists a smooth positive function $c(s)$ of $s$ such that 
$$
(c(s)z, \hat{h}(c(s)z, s))=c(s)(z, h(z, s)), \quad c(s_0)=1
$$
holds. 
Hence we have
\begin{align}
	& \hat{h}(c(s)z, s) =c(s)h(z, s),
\\
	& c'(s)\hat{h}_z(c(s)z, s)+\hat{h}_s(c(s)z, s)=c'(s)h(z, s)+c(s)h_s(z, s).
\label{eqn2}
\end{align}
Since $c(s_0)=1$, Eq.~(\ref{eqn2}) gives
\begin{equation}\label{eqn3}
c'(s_0)\hat{h}_z(z, s_0)+\hat{h}_s(z, s_0)=c'(s_0)h(z, s_0)+h_s(z, s_0),
\end{equation}
that is
\begin{equation}\label{eqn3b}
c'(s_0)(h_0)_z+\hat{h}_s(z, s_0)=c'(s_0)h_0+h_s(z, s_0).
\end{equation}
Differentiating Eq.~\eqref{eqn3b} with respect to $z$ and setting $z=0$, we have 
\begin{equation}\label{eqsz}
\hat{h}_{sz}(0, s_0)=-c'(s_0)h_{zz}(0, s_0)+c'(s_0)h_z(0, s_0)+h_{sz}(0, s_0).
\end{equation}
On the other hand, because $\hat{h}(z, s)=\hat{h}(-z, s)$, we have
\begin{align}
	\hat{h}_s(z, s)=\hat{h}_s(-z, s), \ \hat{h}_{sz}(z, s)=-\hat{h}_{sz}(-z, s),
\end{align}
which imply
\begin{equation}\label{hatz}
	\hat{h}_{sz}(0, s)=0.
\end{equation} 
Assume now that $\lambda_2(s_0)=0$ holds. Then, 
from Eqs.~(\ref{mcd}) and (\ref{hatz}), 
by choosing a suitable eigenfunction $e$ belonging to zero, $\hat{h}_s(z, s_0)$ and $e$ satisfy the same second order ODE, and their values and their first derivatives at $z=0$ coincide.  Hence, by the uniqueness of solutions of second order ODE, they coincide for all $z$, and hence $\hat{h}_s(z, s_0)$ is an eigenfunction belonging to zero. This implies 
\begin{equation}\label{hatzz}
\hat{h}_{sz}(z_2, s)=0.
\end{equation}

Similarly, from $h(z, s)=h(-z, s)$ and $h(z_2+z, s)=h(z_2-z, s)$, we have
$$
h_{z}(0, s)=0, \ h_{sz}(0, s)=0, \ h_{z}(z_2, s)=0, \ h_{sz}(z_2, s)=0.
$$
These facts with (\ref{eqsz}) give
\begin{equation}\label{eqsz2}
0=-c'(s_0)h_{zz}(0, s_0).
\end{equation}
However, $h_{zz}(0, s_0)=0$ does not occur, which will be proved at the end of this section (Lemma \ref{hzz}). Hence $c'(s_0)=0$ holds. 
In this case, from Eq.~(\ref{eqn3b}), we have
\begin{equation}\label{so2}
\hat{h}_{s}(z, s_0)=h_{s}(z, s_0).
\end{equation}
Therefore, $h_s(z, s_0)$ is an eigenfunction belonging to zero, and hence
 $H'(s_0)=0$ holds. 
\hfill$\Box$

\begin{lemma}\label{hzz}
Assume that $X(s)$ satisfies the same assumptions as in Proposition \ref{l2}. Then, $h_{z}(0, s)= 0$ and  $h_{zz}(0, s)\ne 0$ hold.  
\end{lemma}

\noindent {\it Proof.} \ 
We assume that $z=0$ corresponds to a bulge of $X(s)$. In the case where $z=0$ corresponds to a neck of $X(s)$, a similar proof works. 

Recall Eq.~(\ref{mc}), which is equivalent to
\be \label{mc3}
	H(h^{n+1})_z
	=-\Bigl(h^n(1+h_z^2)^{-1/2}\Bigr)_z.
\ee
Hence, we have, for an integration constant $a$, 
\be \label{mc4}
	Hh^{n+1}
	=-h^n(1+h_z^2)^{-1/2}+a.
\ee
This $a$ gives a one parameter family of unduloids with mean curvature $H$. Set 
$$h_{\rm max}(s):=h(0, s).
$$
Then, because $h_z=0$ at the bulge, we have
\be \label{mc6}
	a=Hh_{\rm max}^{n+1}+h_{\rm max}^n.
\ee
Note that for the cylinder, 
\be \label{mcc}
	h_{\rm max}({\rm cylinder})
	=
	\frac{-n}{(n+1)H},
\ee
and hence
\be \label{a}
a({\rm cylinder})
=
\frac{n^n}{(n+1)^{n+1} \vert H \vert^n}.
\ee
Regard $a$ a function of $h_{\rm max}$, and differentiate the both sides of (\ref{mc6}) with respect to $h_{\rm max}$ to get
\be \label{difa}
a'=
H(n+1)h_{\rm max}^n+nh_{\rm max}^{n-1}
=
\{(n+1)Hh_{\rm max}+n\}h_{\rm max}^{n-1}.
\ee
Since
$$
h_{\rm max} \le h_{\rm max}({\rm cylinder})=
	\frac{-n}{(n+1)H},
$$
we have
$$
a'\ge 0.
$$
Hence, 
$$
a \le \frac{n^n}{(n+1)^{n+1} \vert H \vert^n},
$$
and the equality holds if and only if the hypersurface is the cylinder. Therefore, $a$ is a strictly-decreasing function in the family of unduloids with the cylinder as the initial surface. 

Next we regard $h_{\rm max}$ as a function of $a$ and differentiate the both sides of (\ref{mc6}) with respect to $a$ to get
\be \label{max}
h_{\rm max}'(a)=(a')^{-1}=\frac{1}{\{(n+1)Hh_{\rm max}+n\}h_{\rm max}^{n-1}}
> 0, 
\ee
and $h_{\rm max}$ is a strictly-decreasing function in the family of unduloids with the cylinder as the initial surface. 
Hence, 
\be \label{max2}
h_{\rm max} < h_{\rm max}({\rm cylinder})=
	\frac{-n}{(n+1)H}.
\ee
However, if $h_{zz}(0, s)= 0$ holds, 
using $h_{z}(0, s)= 0$, 
from (\ref{mc}), we have
\be \label{mc7}
	h_{\rm max}
	=h(0, s)
	=
	\frac{-n}{(n+1)H},
\ee
which contradicts (\ref{max2}).  
Hence, $h_{zz}(0, s)\ne 0$ must hold.
\hfill$\Box$

\section{Computation of geometric quantities}
\label{sec:num}

\subsection{Integral representations of geometric quantities of unduloids}

The equation for $h(z)$ that the mean curvature of hypersurface is constant can be obtained as the Euler-Lagrange equation
\begin{align}
	\frac{d}{dz}\left( \frac{\partial J}{ \partial h_z } \right) - \frac{\partial J}{\partial h} = 0,
\end{align}
with the following Lagrangian,
\begin{align}
	J(z) = \sqrt{ 1+h_z^2 } \; h^n(z)+ H h^{n+1}(z).
\label{lag}
\end{align}
Here, $H$ is a constant representing the mean curvature of the hypersurface. Since Lagrangian~\eqref{lag} does not depend on $z$ explicitly, the following quantity is conserved,
\begin{align}
	J - \frac{ \partial J }{ \partial h_z } h_z =: C = {\rm const.},
\label{EL2}
\end{align}
which is called the Beltrami identity. Substituting \eqref{lag} into \eqref{EL2}, we obtain an equation like the law of conservation of mechanical energy:
\begin{align}
	\left( \frac{dh}{dz} \right)^2+1-\left( \frac{h^n}{C- H h^{n+1}} \right)^2=0.
\label{eq:pot_form1}
\end{align}
Introducing a new variable $w$ by 
\begin{align}
	w &:= - H h,
\label{wh}
\end{align}
we have
\begin{gather}
	 \left( \frac{dw}{dz} \right)^2 + (-H)^2 U(w)=0,
\label{eq:pot_form2}
\\
	 U(w):=1-\left( \frac{w^n}{K+w^{n+1}} \right)^2,
\;\;\;
	K:=(-H)^n C.
\end{gather}

Denote the zeros of $U(w)$ by $w_\pm \; (0<w_-<w_+)$. Then, it is easy to see that
\begin{align}
	\rho = \frac{w_-}{w_+},
\end{align}
where $\rho=1-s$ is given by Eqs.~\eqref{nonuni} and \eqref{nonuni3}.

Using $U(w_\pm)=0$, one can express $w_\pm$ and $K$ as functions of $\rho$,
\begin{align}
	w_+ &= \frac{1-\rho^n}{1-\rho^{n+1}}
	= \frac{ \sum_{m=0}^{n-1} \rho^m }{ \sum_{m=0}^n \rho^m },
\label{wp}
\\
	w_- & = \frac{\rho(1-\rho^n)}{1-\rho^{n+1}}
	= \frac{ \sum_{m=0}^{n-1} \rho^{m+1} }{ \sum_{m=0}^n \rho^m },
\label{wm}
\\
	K &= \frac{(1-\rho)\rho^n (1-\rho^n)^n}{(1-\rho^{n+1})^{n+1}}
	= \frac{ ( \sum_{m=0}^{n-1} \rho^{m+1} )^n }{ ( \sum_{m=0}^n \rho^m )^{n+1} }.
\label{K}
\end{align}
Here, the last expressions in Eqs.~\eqref{wp}--\eqref{K} are convenient to avoid the round-off errors in the numerical estimation for $0 < \rho \ll  1$ and large $n$.

From Eq.~\eqref{eq:pot_form2} with the assumption that $\frac{dw}{dz} \geq 0$, we obtain
\begin{align}
	dz =  \frac{ dw }{ (-H) \sqrt{-U(w)} }.
\label{dz}
\end{align}
Integrating the left-hand (resp.\ right-hand) side this with respect to $z$ (resp.\ $w$) from $z_1$ to $z_2$ (resp.\ from $w_-$ to $w_+$), we obtain 
\begin{align}
	H
	=
	-\frac{1}{L}
	\int_{w_-}^{w_+} \frac{1}{ \sqrt{ -U(w) } } dw,
\label{H_und}
\end{align}
where $L:=z_+-z_-$ is the half period of the unduloid. This is an integral representation of the mean curvature of an unduloid, which is a function of $\rho$ (or equivalently $s$) and $L$. Then, using Eqs.~\eqref{area}, \eqref{volume}, \eqref{wh}, and \eqref{dz}, the integral representations of the area and volume of a half period of an unduloid are obtained as 
\begin{align}
	V
	&=
	\frac{v_{n+1}}{(-H)^{n+2}}
	\int_{w_-}^{w_+} \frac{ w^{n+1} }{ \sqrt{-U(w)} } dw,
\label{V_und}
\\
	A
	&=
	\frac{a_n}{(-H)^{n+1}}
	\int_{w_-}^{w_+} \frac{ w^{n} \sqrt{1-U(w)}   }{ \sqrt{-U(w)} } dw,
\label{A_und}
\end{align}
which are also functions of $\rho$ and $L$.

Without loss of generality, we can fix the interval $L$ ($L=1$ for example) throughout the stability analysis. Thus, we have expressed the mean curvature, volume, and area of an unduloid as integrals essentially depending only on the non-uniformness parameter $\rho$ (or $s$), Eqs.~\eqref{H_und}, \eqref{V_und}, and \eqref{A_und}. Therefore, as argued in Sect.~\ref{sec:notes}, we do not have to solve $H_0(z)={\rm const.}$ with boundary conditions $h_{0z}(z_1)=h_{0z}(z_2)=0$, which needs an iterative integration to satisfy the boundary conditions at the both boundaries. 

\subsection{Manipulation for accurate numerical integration}

What is necessary to obtain $A(s), V(s),$ and $H(s)$, which play essential roles in the stability analysis, is to estimate the integrals numerically in Eqs.~\eqref{H_und}, \eqref{V_und}, and \eqref{A_und} as accurately as possible. Since $U(w)$ vanishes at the both ends of integral range, the following manipulation helps us to estimate the integrals numerically~\cite{arXiv:0811.2305}. Those who are not interested in the numerics do not need to read the rest of this section.

Integrals~\eqref{H_und}, \eqref{V_und}, and \eqref{A_und} can be rewritten as 
\begin{align}
	Y
	=
	\int_{w_-}^{w_+} \frac{\psi_Y(w)}{ \sqrt{-U(w)} } dw,
\label{Y}
\end{align}
where 
\begin{align}
	\psi_Y (w)
	:=
	\begin{cases}
		\displaystyle -\frac{1}{L} & (Y=H) \\
		\displaystyle \frac{ v_{n+1} w^{n+1} }{ (-H)^{n+2} }  & (Y=V) \\
		\displaystyle \frac{ a_n w^{n} \sqrt{1-U(w)} }{ (-H)^{n+1}   }  & (Y=A) \\
	\end{cases}.
\label{psi_Y}
\end{align}
In order to extract the poles of the integrand, $w_\pm$,  we rewrite the integral as 
\begin{align}
	Y =
	\int_{w_-}^{w_+}
	\frac{ (w^{n+1}+K) \psi_Y(w)  }{ \sqrt{ (w^n + w^{n+1}+K )(w_+-w)(w-w_-)g(w) } } dw,
\label{Y2}
\end{align}
by defining
\begin{align}
	g(w) := \frac{ w^n - w^{n+1} - K }{ (w_+-w)(w-w_-) } = \sum_{p=0}^{n-1} g_p w^p.
\label{gw}
\end{align}
Here, the right-hand side defines the polynomial expression of $g(w)$, which is regular at $w=w_\pm$. The comparison of coefficients yields the following recursion relation and ``boundary conditions'' to be satisfied by $g_p \; (0 \leq p \leq n-1)$, 
\begin{align}
	g_{n-1}&=1,
\\
	(w_++w_-)g_{n-1}-g_{n-2}&=1,
\\
	w_+w_- g_{p+2}-( w_++w_- )g_{p+1}+g_p &=0, \;\;\; (0 \leq p \leq n-3),
\\
	w_+w_- g_1 - (w_++w_-)g_0 &=0,
\\
	 w_+w_- g_0&=K.
\end{align}
These can be easily solved to give the following expression of general term,
\begin{align}
	g_p
	&= (1-\rho^{p+1})(1-\rho^n)^{n-2-p}
	\left( \frac{ \rho }{ 1-\rho^{n+1} } \right)^{n-1-p}
\\
	&=
	\frac{  \left( \sum_{m=0}^p \rho^{m+1} \right) \left( \sum_{\ell=0}^{n-1} \rho^{\ell+1} \right)^{n-2-p} }{ \left( \sum_{k=0}^{n} \rho^k \right)^{n-1-p} },
	\;\;\;
	( 0 \leq p \leq n-1 ).
\label{gp}
\end{align}
Again, the final expression is for the avoidance of round-off error.

Finally, we fix the integration range as 
\begin{align}
	Y
	=
	\int_{-1}^{1}
	\frac{ (w^{n+1}+K) \psi_Y(w)  }{  \sqrt{ (w^n + w^{n+1}+K ) (1-\zeta^2) g(w) } } d\zeta,
\label{Y3}
\end{align}
by changing variable from $w$ to $\zeta$,
\begin{align}
	w= \frac{w_++w_-}{2} + \frac{w_+-w_-}{2} \zeta.
\label{wzeta}
\end{align}

One can accurately estimate $H$, $V$, and $A$ numerically for given $s =1-\rho \in (0,1)$ (after fixing $L$, $L=1$ for example) using Eq.~\eqref{Y3} with Eqs.~\eqref{wp}--\eqref{K}, \eqref{psi_Y}, \eqref{gw}, \eqref{gp}, and \eqref{wzeta}.

\subsection{Area-volume diagrams}

We describe here how to draw the area-volume diagrams in Fig.~\ref{fig:diagrams}. 

First, let us normalize the volume $V$ by the volume of the largest hemisphere, which has a radius identical to the length of the interval $L:=z_2-z_1$, and normalize the surface area $A$ by the surface area of the hemisphere whose radius is $R \in (0,L]$,
\begin{align}
	\hat{V} &:= \frac{ V }{ \frac12 v_{n+2}L^{n+2} },
\\
	\hat{A} &:= \frac{ A }{ \frac12 a_{n+1} R^{n+1}  }.
\end{align}
Here, $V$ and $A$ are the volume and surface area, respectively, of a hemisphere, cylinder, or half-period of unduloid. 

For the hemisphere with radius $R \in (0,L]$, the normalized volume and area are
\begin{align}
	\hat{V}_{\rm hem}
	&=
	\frac{ \frac12 v_{n+2} R^{n+2} }{ \frac12 v_{n+2}L^{n+2} }
	=
	\left( \frac{R}{L} \right)^{n+2},
\;\;\;
	(0 \leq R \leq L),
\\
	\hat{A}_{\rm hem} &= 1,
\end{align}
respectively. These give a parametric representation of the area-volume curve of hemisphere in Fig.~\ref{fig:diagrams} with $R/L \in (0,1]$ being the parameter.

For the cylinder with radius $r \in (0,+\infty)$, the normalized volume and area are given by
\begin{align}
	\hat{V}_{\rm cyl}
	&=
	\frac{ v_{n+1} r^{n+1} L }{ \frac12 v_{n+2}L^{n+2} }
	=
	\frac{ 2 v_{n+1} }{ v_{n+2} } \left( \frac{r}{L} \right)^{n+1},
\;\;\;
	(0 \leq r < +\infty),
\label{Vhat_cyl}
\\
	\hat{A}_{\rm cyl}
	&=
	\frac{ a_n r^n L }{ \frac12 a_{n+1} R^{n+1}  }
	=
	\frac{ 2a_n }{ a_{n+1}  }\left( \frac{r}{L} \right)^n \left( \frac{L}{R} \right)^{n+1},
\label{Ahat_cyl}
\end{align}
respectively. Solving $ \hat{V}_{\rm hem } = \hat{V}_{\rm cyl} $ for $R$, and then substituting it into the right-hand side of Eq.~\eqref{Ahat_cyl}, one obtains
\begin{align}
	\hat{A}_{\rm cyl}
	=
	\left( \frac{2a_n}{a_{n+1}} \right)^{ \frac{1}{n+2} }
	\left( \frac{n+1}{n+2} \right)^{ \frac{n+1}{n+2} }
	\left( \frac{L}{r} \right)^{\frac{1}{n+2}}.
\label{Ahat_cyl2}
\end{align}
Equations \eqref{Vhat_cyl} and \eqref{Ahat_cyl2} give a parametric representation of the area-volume curve of cylinder in Fig.~\ref{fig:diagrams} with  $r/L \in (0,+\infty)$ being the parameter. 

Denoting the volume and area of a half period of unduloid in the form of Eq.~\eqref{Y3} by $V(s)$ and $A(s)$, the normalized volume and area of unduloid are given by
\begin{align}
	\hat{V}_{\rm und}
	&=
	\frac{ V (s) }{ \frac12 v_{n+2}L^{n+2} },
\label{Vhat_und}
\\
	\hat{A}_{\rm und}
	&=
	\frac{ A (s) }{ \frac12 a_{n+1} R^{n+1}  }.
\label{Ahat_und}
\end{align}

Solving $ \hat{V}_{\rm hem} = \hat{V}_{\rm und} $ for $R$ to obtain $R=L( \hat{V}_{\rm und} )^{\frac{1}{n+2}}$, and substituting this into the right-hand side of Eq.~\eqref{Ahat_und}, one obtains
\begin{align}
	\hat{A}_{\rm und}
	=
	\frac{A(s)}{  \frac12 a_{n+1} R^{n+1} \vert_{ R=L( \hat{V}_{\rm und} )^{\frac{1}{n+2}} } }.
\label{Ahat_und2}
\end{align}
Equations \eqref{Vhat_und} and \eqref{Ahat_und2} give a parametric representation of the area-volume curve of unduloids in Fig.~\ref{fig:diagrams} with $s \in (0,1)$ being the parameter.



\end{document}